\documentclass[fleqn,usenatbib]{mnras}
\usepackage{newtxtext,newtxmath}
\usepackage[T1]{fontenc}
\DeclareRobustCommand{\VAN}[3]{#2}
\let\VANthebibliography\thebibliography
\def\thebibliography{\DeclareRobustCommand{\VAN}[3]{##3}\VANthebibliography}

\usepackage{graphicx}	
\usepackage{amsmath}	
\usepackage{pdflscape}

\newcommand{\privcomtBusch}{Busch, M. W. (personal communication, 16 November 2024)}
\newcommand{\privcomtMarshall}{Marshall, S. E. (personal communication, 19 November 2024)}

\newcommand{\DP}{DP$_{\rm 14}$}
\newcommand{\HW}{HW$_{\rm 1}$}
\newcommand{\JD}{JD$_{\rm 6}$}
\newcommand{\JV}{JV$_{\rm 6}$}

\title[Shape of contact binary NEA 2006 DP$_{14}$]{Shape and spin state model of contact binary (388188) 2006 \DP~using combined radar and optical observations}

\author[R. E. Cannon et al.]{Richard E. Cannon$^{1}$\thanks{{\bf E-mail:} richard.cannon@ed.ac.uk {\bf \& ORCID:} 0009-0007-5946-8731},
Agata Ro\.zek$^{1}$,
Marina Brozovi\'{c}$^{2}$,
Petr Pravec$^{3}$,
Colin Snodgrass$^{1}$,
Michael W. Busch$^{4}$,
\newauthor
James E. Robinson$^{1}$,
Abbie Donaldson$^{1}$,
Tanja Holc$^{1}$,
Lance A. M. Benner$^{2}$, 
Shantanu Naidu$^{2}$, 
Peter Ku\v{s}nir\'ak$^{3}$, 
\newauthor
Daniel Gardener$^{1}$, 
Hana Ku\v{c}\'akov\'a$^{3,5,6}$, 
Elahe Khalouei$^{7}$, 
Joseph Pollock$^{8}$\thanks{Deceased}, 
Mariangela Bonavita$^{1}$, 
Petr Fatka$^{3}$, 
\newauthor
Kamil Hornoch$^{3}$,  
Sedighe Sajadian$^{9}$, 
Lara Alegre$^{1,10}$,
Flavia Amadio$^{11}$,
Michael I. Andersen$^{11}$,
Valerio Bozza$^{12,13}$,
\newauthor
Martin J. Burgdorf$^{14}$,
Gabriele Columba$^{12}$,
Martin Dominik$^{15}$,
R. Figuera Jaimes$^{15,16,17}$,
Tobias C. Hinse$^{18}$,
\newauthor
Markus Hundertmark$^{19}$,
Uffe G. J{\o}rgensen$^{20}$,
Penelope Longa-Pe{\~n}a$^{21}$,
Nuno Peixinho$^{21}$,
Markus Rabus$^{22}$,
\newauthor
Sohrab Rahvar$^{9}$,
Paolo Rota$^{12,13}$,
Jesper Skottfelt$^{23}$,
John Southworth$^{24}$,
Jeremy Tregloan-Reed$^{25}$
\\
$^{1}$Institute for Astronomy, University of Edinburgh, Royal Observatory, Edinburgh, EH9 3HJ, UK
$^{2}$Jet Propulsion Laboratory, California Institute of Technology, \\ Pasadena, CA 91109, USA
$^{3}$Astronomical Institute of the Czech Academy of Sciences, Fri\v{c}ova 298, CZ-25165 Ond\v{r}ejov, Czech Republic 
$^{4}$SETI Institute, \\ Mountain View CA 94043, USA
$^{5}$Astronomical Institute, Faculty of Mathematics and Physics, Charles University Prague, V Holešovičkách 2, CZ-180 00 Praha 8,\\ Czech Republic
$^{6}$Research Centre for Theoretical Physics and Astrophysics, Institute of Physics, Silesian University in Opava, Bezručovo nám. 13, CZ-746 01 Opava,\\ Czech Republic
$^{7}$Seoul National University, Gwanak-gu, Seoul, South Korea
$^{8}$Physics and Astronomy Department, Appalachian State University, Boone, NC 28608, USA \\
$^{9}$Perimeter Institute for Theoretical Physics, Waterloo, ON N2L 2Y5, Canada
{$^{10}$Centre for Astrophysics Research, Department of Physics, Astronomy and Mathematics,}\\ { University of Hertfordshire, College Lane, Hatfield AL10 9AB, UK}
$^{11}$Niels Bohr Institute, Jagtvej 155A, 2200 Copenhagen N, Denmark\\
$^{12}$Alma Mater Studiorum - University of Bologna, Dipartimento di Fisica e Astronomia “Augusto Righi”, Via Gobetti 93/2, 40129 Bologna, \\ Italy
$^{13}$Istituto Nazionale di Fisica Nucleare, Sezione di Napoli, Napoli, Italy
$^{14}$Universit{\"a}t Hamburg, Faculty of Mathematics, Informatics and Natural Sciences, \\ Department of Earth Sciences, Meteorological Institute, Bundesstra\ss{}e 55, 20146 Hamburg, Germany
$^{15}$Centre for Exoplanet Science, SUPA,  \\School of Physics \& Astronomy, University of St Andrews, North Haugh, St Andrews KY16 9SS, UK
$^{16}$Millennium Institute of Astrophysics MAS, \\ Nuncio Monsenor Sotero Sanz 100, Of. 104, Providencia, Santiago, Chile 
$^{17}$Instituto de Astrofísica, Facultad de Física, Pontificia Universidad Católica de Chile,\\ Av. Vicuña Mackenna 4860, 7820436, Macul, Santiago, Chile 
$^{18}$University of Southern Denmark, Department of Physics, Chemistry and Pharmacy, SDU-Galaxy,\\ Campusvej 55, 5230, Odense M, Denmark 
$^{19}$Astronomisches Rechen-Institut, Zentrum f{\"u}r Astronomie der Universit{\"a}t Heidelberg (ZAH), 69120 Heidelberg, Germany \\
$^{20}$Centro de Astronom{\'{\i}}a, Universidad de Antofagasta, Av.\ Angamos 601, Antofagasta, Chile
$^{21}$Instituto de Astrofísica e Ciências do Espaço, Departamento de Física, \\ Universidade de Coimbra, 3040-004, Coimbra, Portugal
$^{22}$Departamento de Matemática y Física Aplicadas, Facultad de Ingeniería, Universidad Católica \\ de la Santísima Concepción, Alonso de Rivera 2850, Concepción, Chile
$^{23}$Centre for Electronic Imaging, Department of Physical Sciences, The Open University,\\  Milton Keynes, MK7 6AA, UK
$^{24}$Astrophysics Group, Keele University, Staffordshire, ST5 5BG, UK
$^{25}$Instituto de Astronomia y Ciencias Planetarias, \\ Universidad de Atacama, Copayapu 485, Copiapo, Chile
}
\date{Accepted XXX. Received YYY; in original form ZZZ}

\pubyear{2024}

\begin{document}
\label{firstpage}
\pagerange{\pageref{firstpage}--\pageref{lastpage}}
\maketitle

\begin{abstract}
\noindent
Contact binaries are found throughout the solar system. The recent discovery of Selam, the satellite of MBA (152830) Dinkinesh, by the NASA LUCY mission has made it clear that the term `contact binary' covers a variety of different types of bi-modal mass distributions and formation mechanisms. Only by modelling more contact binaries can this population be properly understood. We determined a spin state and shape model for the Apollo group contact binary asteroid (388188) 2006 \DP~using ground-based optical and radar observations collected between 2014 and 2023. Radar delay-Doppler images and continuous wave spectra were collected over two days in February 2014, while 16 lightcurves in the Cousins R and SDSS-r filters were collected in 2014, 2022 and 2023. We modelled the spin state using convex inversion before using the SHAPE modelling software to include the radar observations in modelling concavities and the distinctive neck structure connecting the two lobes. We find a spin state with a period of $(5.7860\pm0.0001)$ hours and pole solution of $\lambda = (180\pm121)^\circ$ and $\beta = (-80\pm7)^\circ$ with morphology indicating a $520$ m long bi-lobed shape. The model's asymmetrical bi-modal mass distribution resembles other small NEA contact binaries such as (85990) 1999 \JV~or (8567) 1996 \HW, which also feature a smaller `head' attached to a larger `body'. The final model features a crater on the larger lobe, similar to several other modelled contact binaries. The model's resolution is 25 m, comparable to that of the radar images used.

\end{abstract}




\section{Introduction} \label{sec:introduction}

Contact binaries are bi-lobed objects that appear throughout the solar system in both asteroid and comet populations. Notable examples of contact binaries include (25143) Itokawa, a near-Earth asteroid (NEA) visited by the JAXA Hayabusa mission \citep{DemuraEtAl2006_Itokawa-Shape}; Selam, which orbits the main belt asteroid (MBA) (152830) Dinkinesh and was imaged by the NASA LUCY mission in 2023 \citep{LevisonEtAl2024_LUCY-Dinkinesh-Characterisation-Selam-first-look}; and (486958) Arrokoth, a Kuiper Belt object (KBO) imaged by the NASA New Horizons mission in 2019 \citep{PorterEtAl2024_ConfAbs_ArrokothUpdatedShape}. It is estimated from radar observations that at least 15-30\% of NEAs $>200$ m in diameter are contact binaries \citep{BennerEtAl2015A4_15-percent-CB-NEO, VirkkiEtAl2022_30percent-CB}. Furthermore, optical observations suggest that up to 40\%-50\% of smaller Plutinos, a family of KBO, are either elongated or bi-lobed in shape \citep{Thirouin&Sheppard2018_40-percent-CB-in-Plutinos, Brunini2023_50-percent-CB-Plutinos-reason} and there may be several large contact binaries with sizes $> 25$ km in the KBO population \citep{Sheppard&Jewitt2004_2001QC298-CB-KBO}.

We modelled NEA asteroid (388188) 2006 \DP, henceforth \DP, to contribute to the growing number of modelled contact binaries. \DP~is an Apollo group asteroid designated as potentially hazardous with an absolute magnitude of $H = (19.0\pm0.5)$. Being an Apollo family asteroid, \DP's orbital semi-major axis is $1.36$ au, and its eccentricity is $e = 0.78$. This causes it to make frequent close approaches with Mercury, Venus, and Earth \citep{DP14_SmallBodyDatabaseLookup}. \DP~was previously observed twice in 2014 with optical telescopes. Observations by \citet{Hicks2014_DP14observations} estimated a rotational period of $5.78 \pm 0.02$ h and found rotationally averaged colours consistent with an X or C-type spectral classification. \citet{Warner2014_DP14-data-plus-period-estimation} also observed \DP~in 2014, finding lightcurve amplitudes of $1.05$ magnitudes and estimating the period to be $5.77 \pm 0.01$ hours. The large lightcurve amplitudes suggested an elongated object, which, in conjunction with the radar imaging from 2014 (described in detail in Section \ref{sec:OBSERVATIONS}), indicated that \DP~was a contact binary. While no pole estimate was made from either observation, a Yarkovsky detection of $A_2 =\left(-39.508 \pm 6.605\right)\times10^{-15}$ \citep{DP14_SmallBodyDatabaseLookup} implies a rotational pole in the southern hemisphere relative to the plane of the ecliptic.

With the addition of \DP~presented in this work, 23 contact binaries have now been either shape-modelled with ground-based radar and optical observations or imaged directly by spacecraft. Of these, 16 are NEAs modelled primarily or partly with radar observations. In this paper, we shall present the data we collected to create the shape model of \DP~in Section \ref{sec:OBSERVATIONS}, the modelling techniques Section \ref{sec:Modelling} and the results of the modelling in Section \ref{sec:Discussion}, placing them in the context of other contact binaries. 

\section{Observations} \label{sec:OBSERVATIONS}

The orbit of \DP~is such that it is frequently observable from Earth with optical facilities. Since its discovery in early 2006, it has approached within 0.5 au of Earth in 2007, 2014, 2015, 2022, and 2023. \DP~will next be observable in 2030 and 2031. Due to its small size (extending $\sim520$ m in its longest axis), these encounters with Earth are the only opportunities that currently available telescopes can collect data. We used data from 3 epochs in our modelling: 2014, 2022, and 2023. Radar observations were only possible in 2014 as other encounters did not come close enough for current radar observatories to observe (the next close encounter, $<0.05$ au, sufficiently close for a Goldstone-equivalent radar facility to collect continuous wave observations will be in 2065).

\subsection{Optical Lightcurves} \label{subsec:opticaldata}

\begin{table*}
    \centering
    \caption{A list of all \DP~optical lightcurve observations. The site MPC codes are 807 -- Cerro Tololo Inter-American Observatory (CTIO), Chile; 323 -- Perth Observatory, Australia; U82 -- Palmer Divide Station, USA; W74 -- 1.54m Danish Telescope, La Silla, Chile; 950 -- Isaac Newton Telescope, La Palma, Spain. Lightcurve IDs 3 and 4 were taken from ALCDEF \citep{WarnerEtAl2011_ALCDEFconcept}, having been collected by Brian Warner \citep{Warner2014_DP14-data-plus-period-estimation}. Provided are the length of time between the first and final data point in each lightcurve (note that IDs 5-13 are semi-sparse, so do not have uniform coverage over this time span); the distance from the observe to the target, $\Delta$; the solar phase angle $\alpha$ which is the angle made by following a line Sun-Target-Observer; the observer centred ecliptic longitude, $\lambda$ and latitude $\beta$; the filter that was used (IDs 3 \& 4 were taken unfiltered, and then adjusted to be in Johnson V); the exposure time of each observation in seconds (where available, before any stacking of images); and whether they were used in either the Lightcurve or Radar+Lc models. All lightcurves were used to test the final model fits and the initial radar ellipsoid pole scan. Only lightcurves marked for the Radar+Lc model were used for the bi-ellipsoid and vertex fitting, while the spin state was kept constant.}
    \label{tab:AllLightcurves}
    \begin{tabular}{ccccccccccccc} \hline \hline
        ID & Date & Site & Length & $\Delta$ & $\alpha$ & $\lambda$ & $\beta$ & Filt. & Exp. & Lc & Radar+Lc & \\ 
        ~  & ~ & ~ & [h] & [au] & [\textdegree] & [\textdegree] & [\textdegree] & ~ & [s] & Model & Model & \\ \hline
		1  & 2014-02-18 & 807 & 5.3 & 0.116 & 29.4 & 121.6 & -18.0 & R &    & $\bullet$ & $\bullet$ & \\
		2  & 2014-02-19 & 323 & 5.5 & 0.140 & 29.0 & 122.1 & -16.7 & R &    & $\bullet$ & $\bullet$ & \\
		3  & 2014-02-22 & U82 & 5.5 & 0.182 & 29.0 & 122.7 & -15.2 & V &    & $\bullet$ & $\bullet$ & \\
		4  & 2014-02-23 & U82 & 5.3 & 0.198 & 29.1 & 122.8 & -14.8 & V &    & $\bullet$ & $\bullet$ & \\
		5  & 2022-02-23 & W74 & 2.6 & 0.266 & 13.3 & 147.2 & -15.2 & R & 40 & $\bullet$ &   & \\
		6  & 2022-02-25 & W74 & 2.3 & 0.290 & 14.5 & 145.0 & -14.9 & R & 40 & $\bullet$ &   & \\
		7  & 2022-02-26 & W74 & 3.1 & 0.305 & 15.2 & 143.9 & -14.7 & R & 45 & $\bullet$ &   & \\
		8  & 2022-03-02 & W74 & 3.0 & 0.363 & 18.4 & 140.6 & -14.0 & R & 70 & $\bullet$ &   & \\
		9  & 2022-03-04 & W74 & 1.7 & 0.392 & 19.8 & 139.3 & -13.7 & R & 85 & $\bullet$ &   & \\
		10 & 2022-03-06 & 950 & 4.9 & 0.437 & 21.8 & 137.9 & -13.3 & r & 60 & $\bullet$ &   & \\
		11 & 2022-03-07 & 950 & 5.5 & 0.453 & 22.4 & 137.5 & -13.2 & r & 60 & $\bullet$ &   & \\
		12 & 2022-03-08 & 950 & 4.8 & 0.469 & 23.0 & 137.1 & -13.1 & r & 60 & $\bullet$ &   & \\
		13 & 2022-03-09 & 950 & 3.7 & 0.484 & 23.5 & 136.8 & -12.9 & r & 60 & $\bullet$ &   & \\
		14 & 2023-05-26 & W74 & 5.7 & 0.304 & 44.1 & 220.6 & -52.0 & R & 8 & $\bullet$ &   & \\
		15 & 2023-05-29 & W74 & 2.8 & 0.284 & 50.3 & 210.2 & -54.3 & R & 5 & $\bullet$ &   & \\
		16 & 2023-06-09 & W74 & 0.6 & 0.248 & 81.8 & 159.7 & -51.0 & R & 8 & $\bullet$ &   & \\ \hline
    \end{tabular}
\end{table*}

\begin{figure}
	\includegraphics[width=\columnwidth]{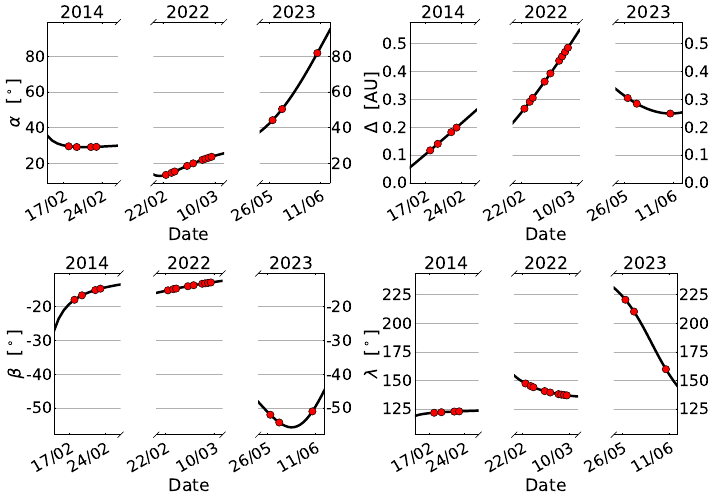}
    \caption{Observing circumstances of the optical observations used in modelling and described in Table~\ref{tab:AllLightcurves}. The black line describes the median value of each property for each night, while the red markers denote the nights that observations took place. \textit{Top Left:} Solar Phase angle $\alpha$, the angle made between Sun-Target-Observer. \textit{Top Right:} $\Delta$, the distance between the observer and the asteroid, in au. \textit{Bottom Left:} $\beta$, Observers' Ecliptic Latitude. \textit{Bottom Right:} $\lambda$, Observers' Ecliptic Longitude.}
    \label{fig:LC_PhaseAnglePlot}
\end{figure}

The lightcurves of \DP~used in this analysis span from February 2014 to June 2023, with the densest coverage in 2022. The phase angle, viewing geometries, and more information for each lightcurve can be seen in Table \ref{tab:AllLightcurves}. The previously published lightcurves (IDs 3 \& 4) and our collected lightcurves demonstrated a two-peaked structure and high amplitudes between 0.5 and 1 magnitudes. The lightcurves used can be seen, in conjunction with the simulated lightcurves of the convex inversion and radar shape models, in Appendices~\ref{figapp:M1_lcfit} and \ref{figapp:rad_lcfit}. We describe our observations from each contributing telescope in the following subsections.

\subsubsection{PROMPT, Cerro Tololo Inter-American Observatory -- 2014}

The Panchromatic Robotic Optical Monitoring and Polarimetry Telescopes (PROMPT) on Cerro Tololo in Chile are owned by the University of North Carolina at Chapel Hill and are $2200$ m above sea level. Consisting of 6 individual 0.41-m telescopes outfitted with Alta U47+ cameras by Apogee with e2v CCDs, the field of view is $10' \times 10'$ with a $1024\times1024$ pixel detector. We observed \DP~for one night with PROMPT in February 2014 in the Cousins R photometric filter \citep{Bessel1990_UBVRI}. Raw image frames were processed using the MIRA software package and reduced using standard photometric procedures. Aperture photometry was performed on the asteroid and three comparison stars.

\subsubsection{R-COP, Perth Observatory -- 2014}

We used the `Remote Telescope Partnership: Clarion University – Science in Motion, Oil Region Astronomical Society, and Perth Observatory' (R-COP) telescope, located in Perth Observatory in Western Australia at an altitude of $386$ metres, to observe \DP~for one night in February 2014 in the Cousins R filter. R-COP's detector has $1600\times1200$ pixels with a $20.2\times15.2$ arcmin$^2$ field of view. Images collected were reduced, and photometry procedures were performed using the same method as the PROMPT telescope.

\subsubsection{Isaac Newton Telescope, La Palma -- 2022} \label{subsec:INTdata}
Between the 6th and 9th of February 2022, we observed \DP~in the SDSS-r filter with the Isaac Newton Telescope (INT). The INT is at an altitude of $2396$ m in the Roque de los Muchachos Observatory on La Palma and is owned by the Isaac Newton Group of Telescopes. We used CCD4 of the Wide-Field Camera (WFC), which has $2000 \times 4000$ pixels covering an $11\times22$ arcmin$^2$ field of view. The images were reduced using standard bias subtraction and flat-fielding methods. Aperture photometry was performed by calculating the average FWHM of each frame using a PSF fitting function on all non-overexposed sources. The resulting FWHM was then used to construct an aperture to measure the flux of \DP~and the background objects. The lightcurve of \DP~was then calibrated to all ATLAS-RefCat2 catalogue stars \citep{TonryEtAl2018_ATLASREFCAT2} found in the list of background sources. Crossmatching of sources to catalogue stars was performed with the calviacat package \citep{Kelley&Lister2019_CalviacatRef}.

\subsubsection{Danish 1.54-metre telescope, La Silla -- 2022 \& 2023}

The 1.54-m Danish Telescope is located at La Silla Observatory, Chile, at an elevation of 2366 m. It is operated jointly by the Niels Bohr Institute, University of Copenhagen, Denmark, and the Astronomical Institute of the Academy of Sciences of the Czech Republic. All images of \DP~were obtained by the Danish Faint Object Spectrograph and Camera (DFOSC) with an e2v CCD 231 sensor and standard Cousins R filter. The CCD sensor has $2048 \times 2048$ square pixels ($13.5~\mu$m size), which were used in the $1 \times 1$ binning mode resulting in a scale of $0.396$ pixels$^{-1}$ and a $13.5 \times 13.5$ arcmin$^2$ field of view.

All images were reduced using standard flat-field and bias-frame correction techniques. For the observations taken in 2022, half-rate tracking was used such that the star and asteroid images present the same trailing in one frame, facilitating robust photometry. The photometry was performed using {\it Aphot}, a synthetic aperture photometry software developed by M.\,Velen and P.\,Pravec at Ond\v{r}ejov Observatory. It reduces asteroid images with respect to a set of field stars, and the reference stars are calibrated in the Johnson–Cousins photometric system using \citet{Landolt1992_StandardStars} standard stars on a night with photometric sky conditions. This resulted in R-magnitude errors of about 0.01 mag. Typically, eight local reference field stars, which were checked for stability (non-variable, not of extreme colours), were used each night.

The 3 nights in 2023 occurred while the target crossed the galactic plane with a very high rate of motion, so the median stack of every $5$ frames was used to reduce the effect of background sources around the target. Photometry of the reduced images was performed in the same way as the INT observations. Due to the high rate of motion of the target, its track over a single night exceeded the field of view of DFOSC; these nights were split into multiple fields and calibrated independently.

\subsubsection{Published Data}
Data collected by \citet{Warner2014_DP14-data-plus-period-estimation} at the Palmer Divide Station in the United States in 2014 were also used. These data are publicly available on the Asteroid Lightcurve Data Exchange Format (ALCDEF) database \citep{WarnerEtAl2011_ALCDEFconcept}.

\subsection{Planetary Radar} \label{subsec:planetaryradar}

\begin{figure}
	\includegraphics[width=\columnwidth]{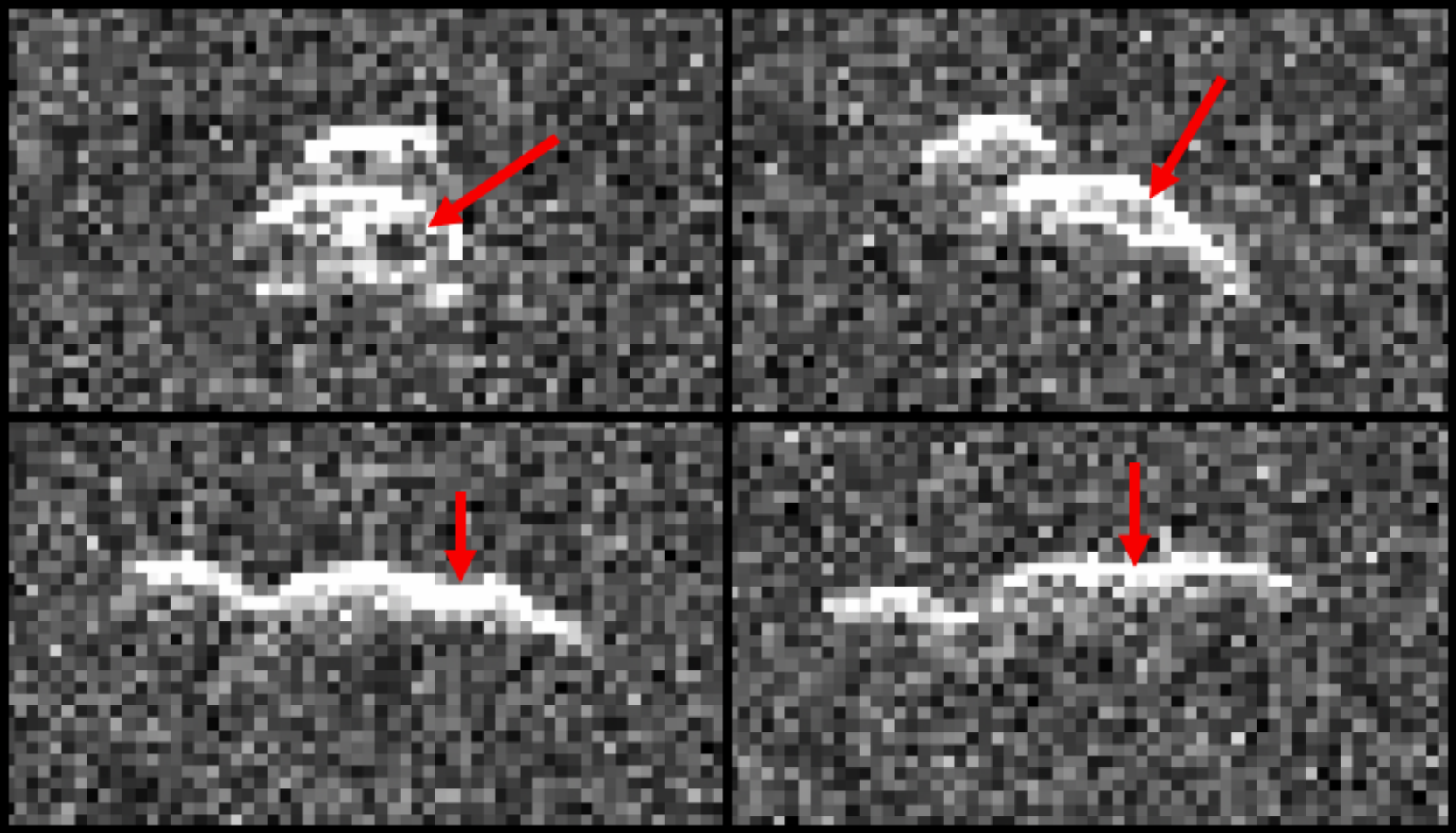}
    \caption{A collage of 4 delay-Doppler images taken on 12/02/2024 from the Goldstone antenna in California, USA. These images contain information on the Doppler-shift on the horizontal axis, increasing from left to right, and the time-delay on the vertical axis, with the delay increasing from top to bottom. Images are $9.6$ Hz by $4.25~\mu$s (or $1.274$ km) large. The red arrows point to features in each image that indicate a crater on the larger lobe, highlighted best by the difference in delay.}
    \label{fig:DP14_dd_image}
\end{figure}

\begin{table*}
    \centering
    \caption{A list of all radar observations used in shape modelling. All observations were collected at the Goldstone DSN 14 antenna. Columns show the date of observations, the start and end time of receiving the signal, the type of observation (CW - continuous wave and DD - delay-Doppler), the baud and code length and frequency resolution, the number of runs and looks per observation, and the orbital solution used. Continuous wave spectra also have the recorded SNR for the OC signal and the ratio between the received SC and OC signals. $\sigma_{OC}$ is the cross-sectional area of the reflected signal, and $A_{\rm proj}$ is the cross-sectional area of the final model at the time of these observations. The albedo is then calculated with $\sigma_{OC} / A_{\rm proj}$. For these observations, the transmitted power, $T_x$ was within 10\% of $430$ kW.}
    \label{tab:AllRadarObs}
    \begin{tabular}{ccccccccccccc} \hline \hline
        Date        & Time                  & Type  & Baud              & Resolution& Runs&Looks& Solution& OC     & $\sigma_{OC}$  & $\rm SC/OC$               & $A_{\rm proj}$& Albedo\\ 
        ~           & ~                     & ~     & [$\mu$s]          & [Hz]      & ~   & ~   & ~       & SNR    & [$\rm km^2$]   &                           & [$\rm km^2$]  &\\ \hline
		2014-02-12  & 03:58:08 - 04:03:41   & CW    & --                & 0.5       & 6   & 8   & 24      & 265.24 & 0.017          & $0.641^{+0.005}_{-0.004}$ & 0.075         & 0.23\\
        2014-02-12  & 04:03:41 - 04:09:14   & CW    & --                & 0.5       & 6   & 8   & 24      & 263.02 & 0.018          & $0.684^{+0.004}_{-0.005}$ & 0.075         & 0.24\\ 
        2014-02-12  & 04:11:01 - 04:15:20   & CW    & --                & 0.5       & 5   & 8   & 26      & 244.58 & 0.019          & $0.638^{+0.005}_{-0.004}$ & 0.065         & 0.29\\ 
        2014-02-12  & 05:02:34 - 07:30:15   & DD    & 0.125             & 0.16      & 153 & 4   & 30      & & &                                                 &          & \\ \hline
        2024-02-13  & 03:49:53 - 03:56:00   & CW    & --                & 0.133     & 5   & 11  & 34      & 46.27  & 0.015          & $0.630^{+0.026}_{-0.024}$ & 0.072         & 0.21\\
        2024-02-13  & 04:27:05 - 06:29:43   & DD    & 0.25              & 0.26      & 88  & 10  & 34      & & &  \\ \hline
        
    \end{tabular}
\end{table*}

As radar observations are performed by emitting a signal towards a target and measuring the reflection, the strength of the signal is proportional to the inverse fourth power of distance. Therefore, near-Earth objects and the largest objects in the main asteroid belt are the predominant targets of radar observations for small solar system objects \citep{DurechEtAl2015A4_AsteroidModelsFromMultipleDataSources}.

Ground-based radar observations of \DP~were performed at the Goldstone Deep Space Network antenna in California, USA. We collected a mix of delay-Doppler imaging and continuous-wave power spectra of \DP~over the 12th and 13th of February in 2014 (Table~\ref{tab:AllRadarObs}) while the target was at distances between $0.02$ and $0.05$ au. Observations consist of transmitting $8560$ MHz (3.5 cm wavelength) radio waves and recording the reflected signal. We transmitted continuous (CW) and binary phase-coded (BPC) waveforms. Each observation contains several `looks', statistically independent measurements of the returning signal, which reduce the signal's noise by a factor of $\sqrt{N_{\rm looks}}$. The reflected echo power spectra obtained via CW carry Doppler-only information about the object's instantaneous line-of-sight velocity, size and rotation properties, and radar scattering properties. Specifically, we measured the ratio of the reflected signal with the same circular polarisation (SC) and the opposite circular polarisation (OC). The delay-Doppler images obtained with BPC contain the Doppler information and the time delay of the signal reflecting back to the observer. The combination of the radial velocity of different parts of the target's surface and the corresponding line-of-sight distance from the time-delay makes delay-Doppler images particularly valuable for shape modelling and size determination. Further explanation of radar observations and the techniques employed is available in \citet{VirkkiEtAl2023_RadarReview} and \citet{MagriEtAl2007_Betualia-Shape-Radar-Explanation}.

Delay-Doppler images, while not equivalent to an optical plane-of-sky view, can be visually inspected to gain insight into the object's shape before modelling begins. The example delay-Doppler frames showing \DP~in Fig.~\ref{fig:DP14_dd_image} have the Doppler-shift (equivalent to the radial velocity) increasing from left to right on the horizontal axis and the time-delay (equivalent to line-of-sight distance) increasing from top to bottom on the vertical axis. All $4$ images demonstrate a clear bi-lobed structure with an unequal mass distribution.

An inspection of the delay-Doppler images can be used to estimate the physical extent and size of the object before modelling begins. This can be done by counting the extent of the signal. Inspection of these delay-Doppler images revealed an unequal mass distribution of an object approximately $490$ m long, with a small spherical lobe of $\sim 112$ m in diameter and a larger elliptical lobe $\sim 224$ m wide and $\sim 262$ m long. The neck was estimated to be only $75$ m in diameter. Additionally, there is evidence of a crater on the larger lobe seen by features predominantly differing in the delay axis, pointed out with red arrows in Fig.~\ref{fig:DP14_dd_image}.

\section{Modelling} \label{sec:Modelling}

Our shape modelling procedure consisted of two parts. First, we constrained the object's spin state based on optical data only to produce a convex shape of \DP. Then, we used this spin state solution to include both optical and radar data to refine the shape to fit the bifurcated appearances portrayed in the delay-Doppler images. When modelling, delay-Doppler images have a north-south ambiguity for rotational pole solution, so complimentary lightcurves are vital to constrain the spin state of the target. All the lightcurves were used when fitting the spin-state of \DP to utilise the different viewing angles of each observation. However, when fitting details of the shape using radar observations, the spin state was kept constant, and only lightcurves from 2014 were used.

\subsection{Convex Inversion} \label{subsec:Convex inversion}

\begin{figure}
	\includegraphics[width=\columnwidth]{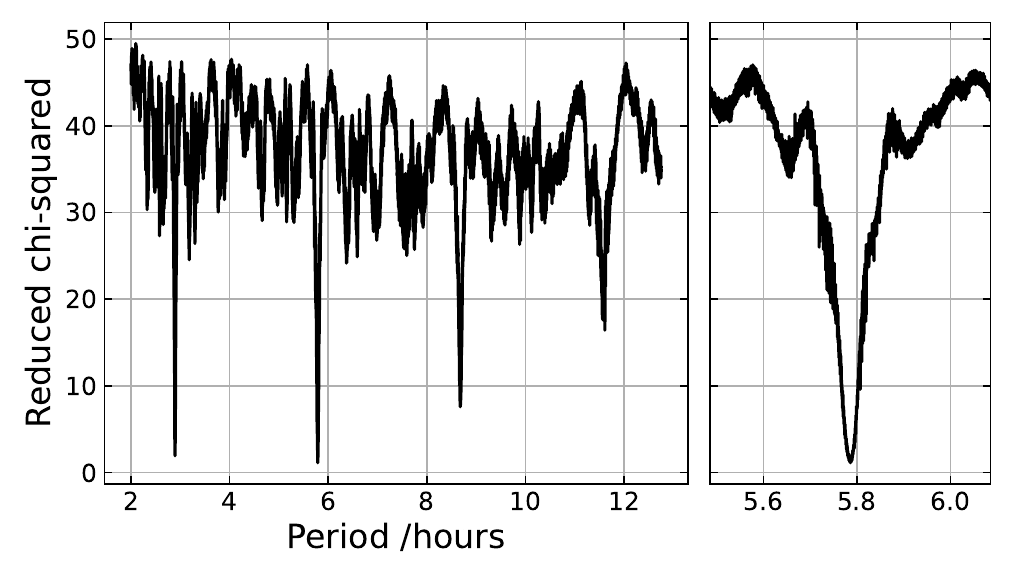}
    \caption{Result of the period scan of \DP~using convex inversion, described in Section \ref{subsec:Convex inversion}. A range of 2 to 13 hours was used, resulting in a best fit of $(5.7860\pm0.0001)$ hours. Significant minima also appear at $1/2$ and $3/2$ multiples of this value but did not result in as good a fit. A cutout to the right shows a zoom-in of the minima.}
    \label{fig:ConvInv_PerScan}
\end{figure}

We used convex inversion \citep{Kaasalainen&Torppa2001_ConvInv1, KaasalainenEtAl2001_ConvInv2} to constrain the spin state of \DP. Following the same procedure as \citet{RozekEtAl2019_JV6-shape}, we modelled six equidistant pole solutions for a range of periods between 2 and 13 hours and recorded the best pole solution for each period. This period range was considered sufficient due to the period assessment of \citet{Warner2014_DP14-data-plus-period-estimation} using just the two published lightcurves (Lightcurve IDs 3 \& 4) of $5.77 \pm 0.01$ hours, and \citet{Hicks2014_DP14observations} independently calculating a value of $5.78 \pm 0.02$ hours. Combining \citet{Warner2014_DP14-data-plus-period-estimation} data and our new lightcurves, we find the best fit at a sidereal period of $5.7860\pm0.0001$. This error is the range of periods for which the reduced $\chi^2$ of the period scan is less than 10\% from the minimum value. Fig.~\ref{fig:ConvInv_PerScan}, displaying the best result for every period, also has significant peaks at 0.5, 1.5 and 2 times the best period solution. This is due to the symmetrical two-peaked nature of the lightcurves created by an elongated object such as a contact binary.

\begin{figure}
	\includegraphics[width=\columnwidth]{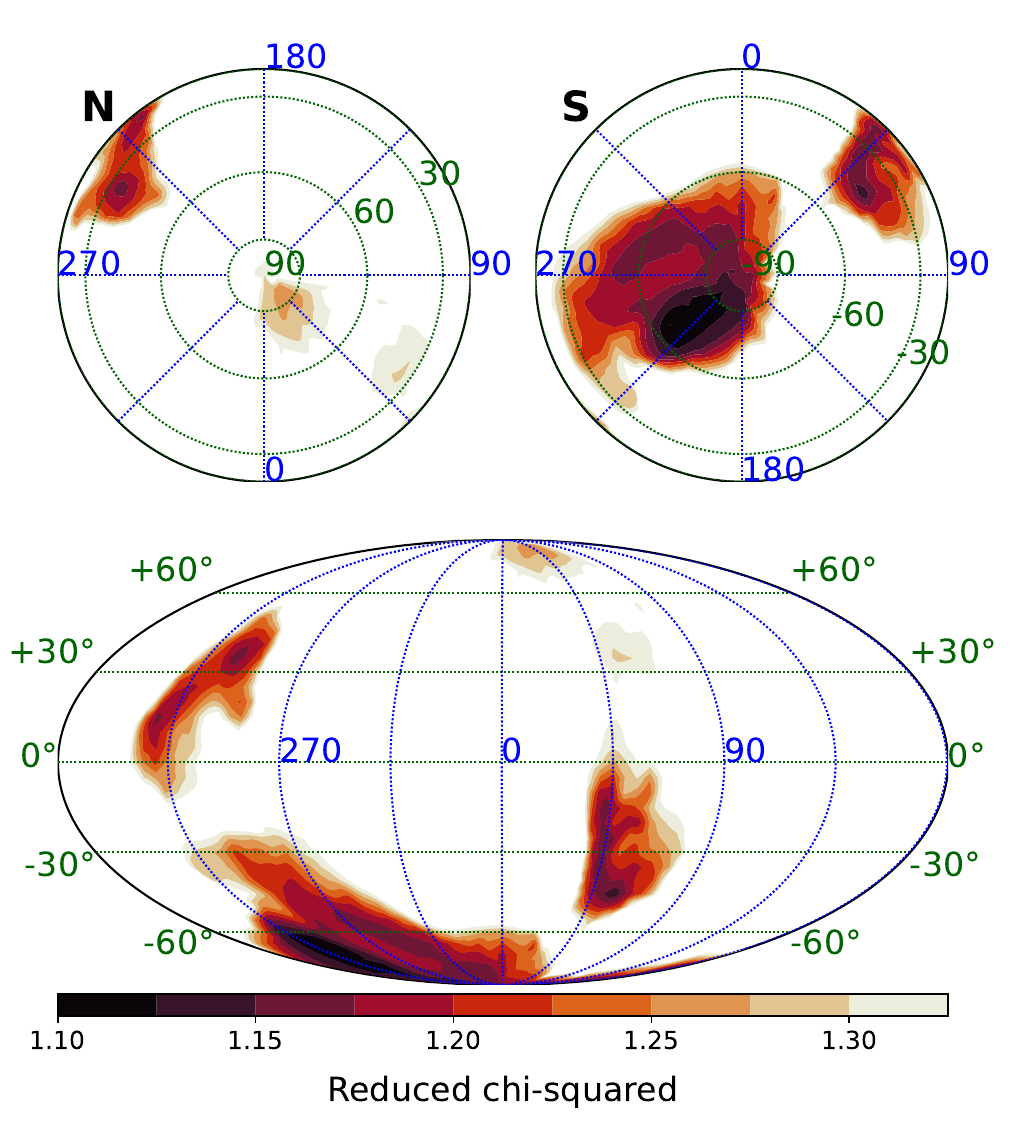}
    \caption{Result of the pole scan of \DP~using convex inversion, described in Section \ref{subsec:Convex inversion}. Results are shown in coordinates relative to the ecliptic plane, with darker areas representing those of a better fit. Any white areas are more than 1.25 times the minimum calculated value. Orthographic projections of North and South are shown above a Mollweide projection. The best result was found at $\lambda = 235^{\circ}$ and $\beta = -65^{\circ}$.}
    \label{fig:ConvInv_PoleScan}
\end{figure}

With the period scan complete, we created a $5^{\circ} \times 5^{\circ}$ grid of pole solutions in ecliptic coordinates. Here, the pole is kept constant while the period (with initial condition input from above) is allowed to vary while the shape model is created. The resulting grid of pole solutions, shown in Fig.~\ref{fig:ConvInv_PoleScan}, indicates two pairs of solutions, one at the two poles, and the other around $\beta = \pm 30^\circ$ and $\lambda = 45^\circ {\rm ~or~} 275^\circ$. These pairs are caused by the ambiguity between mirroring solutions as proven in \citet{KaasalainenLamberg2006_AmbiguityOfMirrorSolutions}. The pole solution of $\lambda = 235^{\circ}$ and $\beta = -65^{\circ}$ had a minimum value of $\chi^2$.

While convex inversion is less useful for contact binary objects due to its inability to model any concavities such as craters or a concave neck structure \citep{Harris&Warner2020_LC_CantTellBrickFromCB}; large, flat surfaces on convex inversion models can indicate the presence of large scale concavities \citep{DevogeleEtAl2015_ConvInvFlatSurfacesConcavities}. The best of these models and the corresponding lightcurve fits are shown in Appendix \ref{app:ConvInvResults}. 
This provides estimates for the aspect ratios of the object and its period, which we used in the initial conditions of the radar modelling.

\subsection{Radar Modelling} \label{subsec:Radar modelling}

We used the SHAPE modelling software \citep{MagriEtAl2007_Betualia-Shape-Radar-Explanation} to integrate the radar observations with the optical lightcurves. 
This powerful tool can accurately create simulated delay-Doppler and CW observations for a model to compare to the observed data. 
When using SHAPE, we again followed the procedure in \citet{RozekEtAl2019_JV6-shape}. 
We first masked the delay-Doppler images and CW spectra to the region only around the signal. 
This is done with a grid of pixels with 0 or 1, which limits the data that SHAPE will attempt to fit to, and in practice, `crop' the images and spectra such that SHAPE does not attempt to fit any noise around the signal. At longer delay values, it can be harder to visually distinguish the signal from noise. We kept a region of $15$ pixels beyond the visible extent of the signal in all frames before masking pixels.
Additionally, points on each of the lightcurves were binned into groups of five by taking the mean of consecutive data points in order to reduce the noise in the data. 
This allows the core information of the spin state to be preserved, but forces SHAPE to model the structure of \DP~using the radar images predominantly rather than attempting to fit small-scale structures from the lightcurves.

\begin{figure}
	\includegraphics[width=\columnwidth]{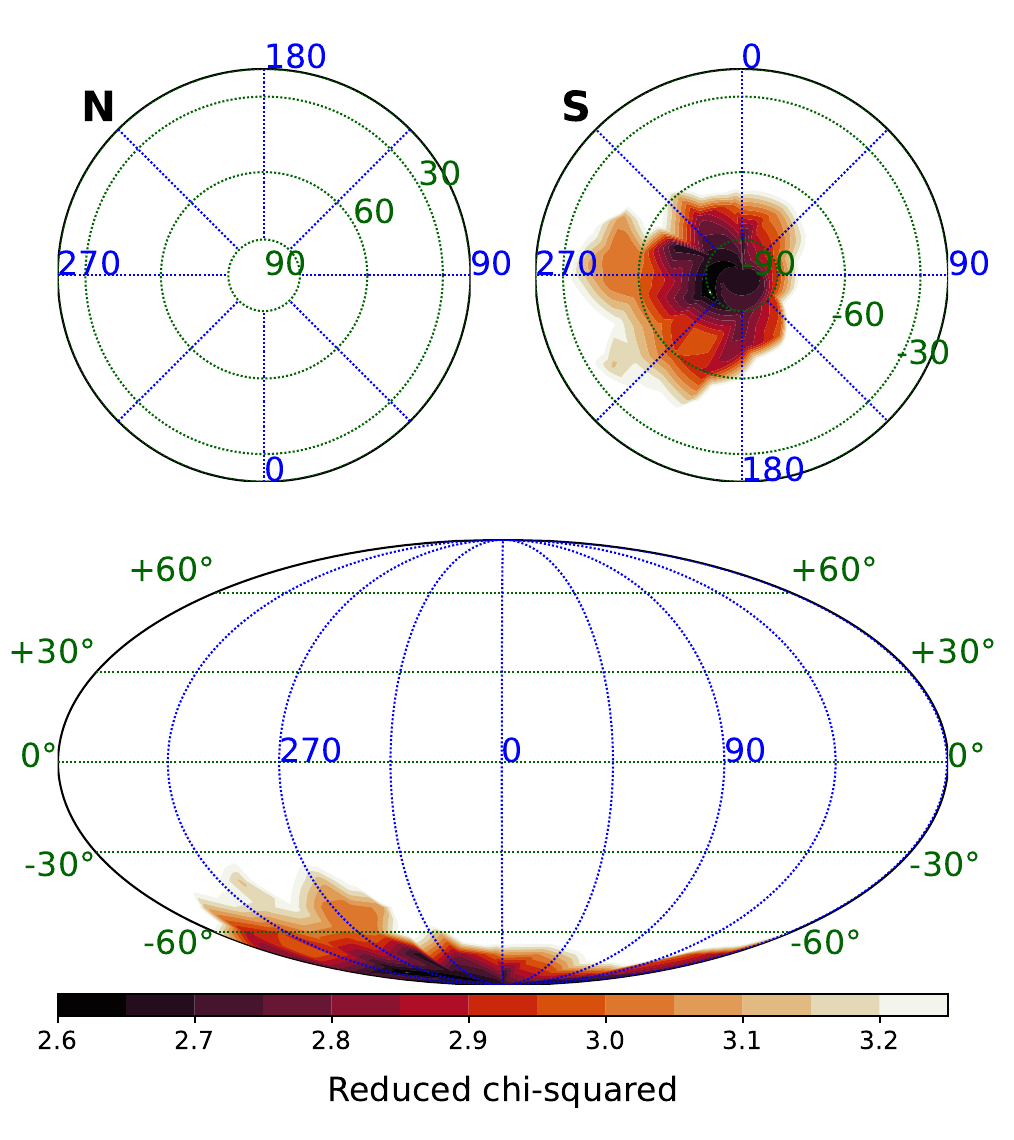}
    \caption{Result of the single-ellipsoid pole scan of \DP~using SHAPE, described in Section \ref{subsec:Radar modelling} using all of the lightcurves and radar data available. The initial conditions of the input model were that of a dynamically equivalent ellipsoid of equal volume (DEEVE) to the best convex inversion solution, but all shape parameters were allowed to vary. Results are shown in coordinates relative to the ecliptic plane, with darker areas representing those of a better fit. Any white areas are more than 1.25 times the minimum calculated value of $\chi^2$. Orthographic projections of North and South are shown above a Mollweide projection.}
    \label{fig:Radar_PolScan}
\end{figure}

The standard method of creating a model with SHAPE is to start with a simple model and build complexity, as SHAPE iterates only one parameter at a time when fitting a model to data. We started with a single ellipsoid model in a fixed grid of $10^{\circ} \times 10^{\circ}$ pole solutions to recreate the convex inversion pole scan results using all of the collected lightcurves and the radar data. Due to the spherical geometry, there is only a small distance between differing longitudinal coordinates, so pole solutions were spaced farther apart in order to use fewer computing resources. Therefore, once the specific convex inversion solutions were added to the pole scan, only 435 pole solutions were tested. The spin state of an object in SHAPE is described with the rotational period, the two pole angles, $\lambda$ and $\beta$, and a rotation phase, $\psi$, for a given epoch that we selected to be midnight before the first observation took place. To perform the pole scan, we first kept the ratios of the ellipsoid constant at the same ratios as the Dynamically Equivalent Equal Volume Ellipsoid (DEEVE) of the best convex inversion solution, fitting only the period, $\psi$, and the length of the ellipsoid ($2a$). By using a very large step size for $\psi$, we ensured that each of the 435 models was orientated with their long axis in line with the delay-Doppler images before we allowed the ratios $a/b$ and $b/c$ for the ellipsoid to vary. The result of the subsequent pole scan with free ellipsoid parameters is in Fig.~\ref{fig:Radar_PolScan}. We found that the previous convex inversion solutions created very good fits to the lightcurve data but were out of phase with the radar delay-Doppler images. This is not the case with the southern pole solutions, despite the ellipsoid producing slightly worse fits to the lightcurve data.

To increase the complexity beyond a single ellipsoid, we proceeded with the pole solutions within 10\% of the best-fit value at $(\lambda,~\beta) = (-80^\circ,~240^\circ)$. Further modelling took place with only the radar data and 2014 optical data to better focus on fitting the morphology of \DP. As the spin state information in the latter lightcurves was removed, the spin state information was fixed at this point in the modelling process, such that it would not alter the period to better fit the 2014 data and worsen the fit to the 2022 and 2023 data. Having first manually created bi-ellipsoid models for each of the solutions, we again allowed SHAPE to vary $\psi$ (keeping the period and pole angles fixed) before fixing all spin state information and varying the size, location and orientation of each ellipsoid. At this point, several solutions could not create contact binary structures with the two ellipsoid components in contact despite the neck being clearly visible in some of the delay-Doppler images collected (see Appendix~\ref{figapp:rad_ddfit1}). Therefore, these solutions were discarded, and modelling continued with only the pole solutions of $\beta = -80^\circ$ and $\lambda = 0^\circ$, and $180^\circ$, and $\beta = -90^\circ$. 

We then created 300-vertex models for the remaining solutions to better model the neck, ensuring that the model's resolution was large enough to fit over the noise in the data, refining them to 500 vertices once the initial vertex fit had been completed. A 500 vertex model produced an average side length of the facets of $\sim25{\rm m}$. As the resolution of the delay-Doppler images only equated to $\sim19{\rm m}$, the resolution was not increased beyond this to reduce the effects of over-fitting noise in the data.

We introduced penalty functions to create vertex models to discourage non-physical solutions. These included discouraging non-principal-axis rotation (a more complex case to model that, as of yet, there is no evidence for) and a smoothness parameter to discourage `spiky' models that occur when a single vertex moves far from the main body to fit a single pixel of noise.

The resulting best-fit vertex model has pole solution $\lambda = 180^\circ$ and $\beta = -80^\circ$. This model was significantly better than the others as it was the only solution with lightcurve amplitudes in the 2023 observation epoch equal to the data. The variation in amplitudes for this epoch can be attributed to the different viewing geometries and high phase angle of observations, where the amount of reflected light would be more highly dependent on small changes in the orientation of the rotational pole due to an increase in the self-shadowing effects of any concavities \citep{KaasalainenEtAl2002_AsteroidModelsFromDiskIntegratedData}. Therefore, we proceeded with only the best model.

Due to the increased uncertainty when modelling the z-extents, the model was stuck in a local minimum, with the larger lobe being flatter rather than more elliptical. Using Blender \citep{Blender}, we manually adjusted this lobe to be closer to an ellipsoid by sculpting additional volume onto the flat surfaces of the model in the $z$ axis. This was done to avoid stretching the model as a whole, which would also affect the smaller lobe. By doing so, we reduced the calculated $\chi^2$ of the model by $10\%$. Further fitting iterations were then performed using a combination of smaller step sizes and smaller tolerances, reducing the penalty functions for concavities. Remarkably, the crater on the larger lobe was clearly modelled, even with penalties still in place to discourage concave solutions. We also performed modelling attempts with more minor penalties to better encourage modelling of the concave neck and crater structures; however, over-fitting of the noise quickly resulted in `spiky' features appearing on the model. The final model was selected as a compromise between over-fitting the noise and replicating the crater to the best of our abilities.

The resulting best-fit vertex model had pole solution $\lambda = (180\pm121)^\circ$ and $\beta = (-80\pm7)^\circ$ with a period of $5.786$ hours, in agreement with the convex inversion solution. These errors in the rotational pole are conservative estimates based on a statistical analysis of the pole solutions selected from the single ellipsoid pole scan. While the error in $\lambda$ appears high, due to the nature of spherical geometry close to the poles, there are only minor differences between differing longitude values. The selected solution was significantly better than the others as it was the only model with lightcurve amplitudes in the 2023 observation epoch matching the data. The model has 500 vertices, with an average side length of $26.5$ m, slightly larger than the spatial resolution of the radar data of $19$ m. Having the side length larger than the spatial resolution reduces the effect of noise on the model. The model is shown in Fig. \ref{fig:Radar_best_model}, and the lightcurve and delay-Doppler fits are in Appendix \ref{app:RadarModellingResults}.

\begin{figure*}
	\includegraphics[width=\textwidth]{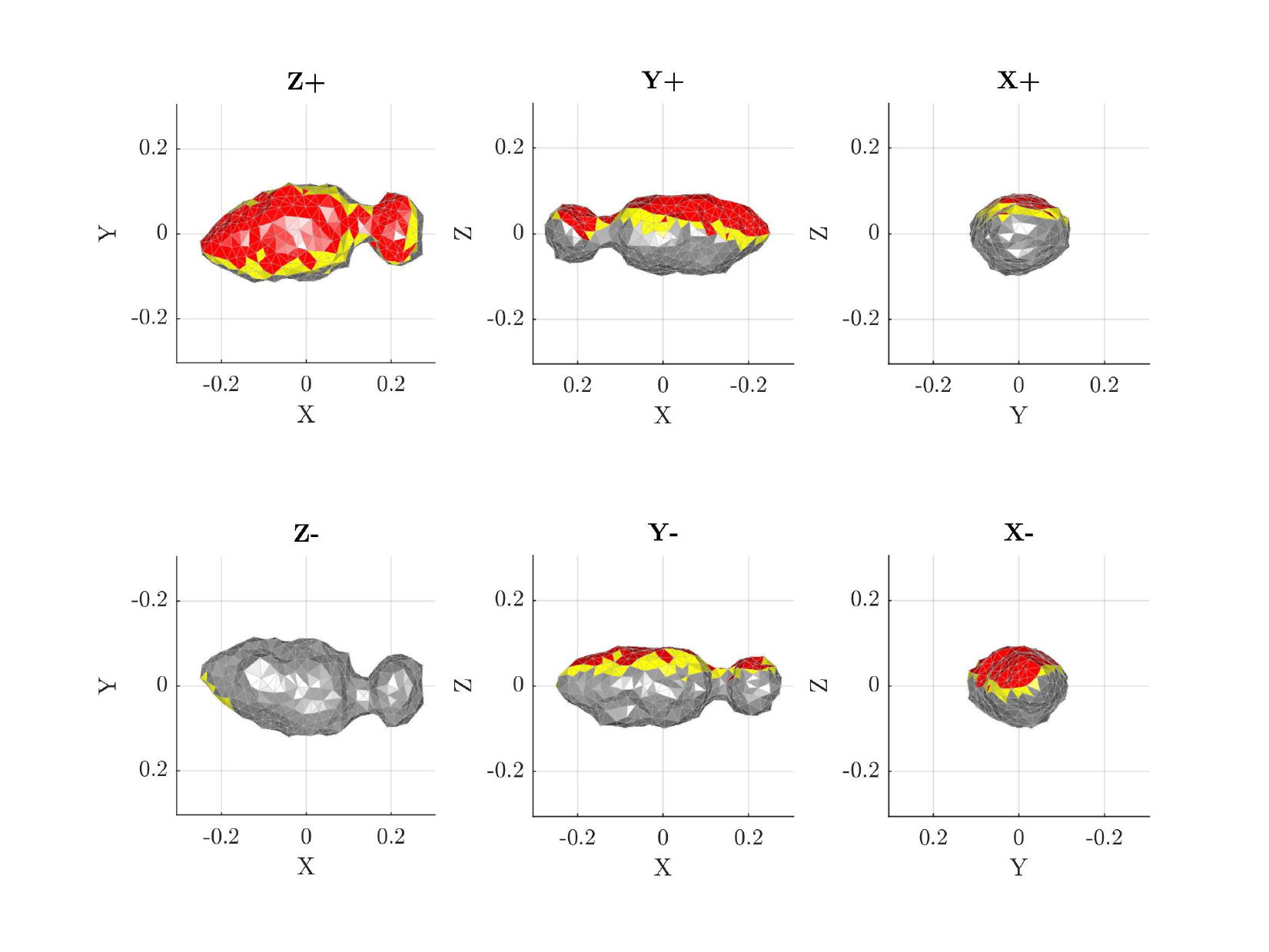}
    \caption{The final shape model of \DP~produced by the method explained in \ref{sec:Modelling}, using a combination of optical and radar observations. The model has pole solution $\lambda = 180$, $\beta = -80$. Red shading is applied to facets not viewed by the delay-Doppler radar imaging, while yellow shading is applied to facets viewed only at scattering angles greater than $60^\circ$. The six plots show the views along the X, Y, and Z axes from both the positive and negative ends and the scales on the axes are in kilometres. The model is $520\pm80$ m long with an average facet length of $25$ metres. The largest fractional uncertainty is along the z-axis of the model.}
    \label{fig:Radar_best_model}
\end{figure*}

\section{Discussion} \label{sec:Discussion}

\subsection{Shape and gravitational environment} \label{subsec:DP14 Shape and gravity}

\begin{table}
    \centering
    \caption{Properties of (388188) 2006 \DP~as derived from the radar+lc shape and spin state model. The uncertainty in the period is derived from a 10\% deviation of reduced $\chi^2$ from the best-fit value in the period scan. Other uncertainties are conservative estimates derived from statistical analysis of the single ellipsoid models within 10\% $\chi^2$ of the best solution of the radar ellipsoid pole scan.}
    \label{tab:388188shapeProperties}
    \begin{tabular}{lcr} \hline \hline 
    Parameter               & Value     & Uncertainty  \\ \hline
    $D_{eq}$ [km]          & 0.262     & 0.037 \\
    Volume [km$^3$]         & 0.009     & 0.005 \\
    Surface Area [km$^2$]   & 0.281     & 0.075 \\ \hline
    Physical Extents [km]   &           &  \\
    ~~~~X                   & 0.527     & 0.080 \\
    ~~~~Y                   & 0.236     & 0.028 \\
    ~~~~Z                   & 0.192     & 0.043 \\
    DEEVE Extents [km] \& Ratios &      &  \\
    ~~~~2a                  & 0.522     & 0.080 \\
    ~~~~2b                  & 0.203     & 0.028 \\
    ~~~~2c                  & 0.171     & 0.043 \\ 
    ~~~~a/b                 & 2.57      & 0.19 \\
    ~~~~b/c                 & 1.19      & 0.37 \\ \hline
    Spin state              &      & \\
    ~~~~P [h]               & 5.7860     & 0.0001 \\
    ~~~~$\lambda~[^\circ]$  & 180       & 121 \\
    ~~~~$\beta~[^\circ]$    & -80       & 7  \\ \hline
    \end{tabular}
\end{table}

The rotational period of $(5.7860\pm0.0001)$ h and rotational pole of $\lambda = (180\pm121)^\circ$ and $\beta = (-80\pm7)^\circ$ are in agreement with the previously found period of $5.78 \pm 0.02$ h by \citet{Hicks2014_DP14observations} and the negative Yarkovsky detection of $\left(-39.508 \pm 6.605\right)\times10^{-15}$ \citep{DP14_SmallBodyDatabaseLookup} which indicated a pole solution in the southern hemisphere. While the convex inversion was critical in refining our estimate for the rotational period, its convex shape approximation does not perfectly replicate a `gift-wrapped' version of the final radar model (Appendix \ref{app:ConvInvResults}). This is likely due to the differing pole solutions used in each case. The rotational pole that produced the best model with convex inversions did not allow any radar model to be in phase with observations while still providing lightcurves of the correct amplitude in the 2023 epoch. Appendix \ref{app:RadarModellingResults} demonstrates a slight offset on some sections of lightcurves 14-16. As the radar model uses the same rotational period (within the given uncertainty) as the convex inversion solution, and the phase offset is inconsistent within the individual lightcurves, this is likely an effect caused by differences in the modelled shape of \DP.

\DP's shape (Fig.~\ref{fig:Radar_best_model}) is a long, thin object consisting of two unequally sized lobes connected with a narrow neck. While just over $520$ metres long, it is only $230$ meters across at its widest point, and the neck connecting the two lobes has an equivalent radius of $39$ meters (defined as the radius of an equal diameter circle corresponding to the smallest cross-sectional area of the final model). Due to the resolution of the delay-Doppler images, the average edge length of the model is $25$ m, limiting our ability to distinguish fine surface features of the smaller lobe and all but the largest of features on the larger lobe. When split along the narrowest cross-sectional area of the neck, the smaller lobe is more spherical, with physical extents of $134$, $220$, and $140$ metres, while the larger lobe is more elongated with extents of $404$, $230$, and $180$ metres in the X, Y, and Z axes respectively (where the X axis is aligned with the long axis of the body).

As Fig.~\ref{fig:Radar_best_model} demonstrates using red shading (and can also be seen in Appendix~\ref{figapp:rad_ddfit1}), there is a significant portion of the body on the southern hemisphere of \DP~which was not imaged with radar due to the rotational poles orientation. As such, this region of the surface is modelled by SHAPE to fit the lightcurves best and maintain reasonable physical properties of mass distribution with respect to the centre of mass. Unfortunately, as discussed in Section \ref{subsec:FutureObservations}, there are very few opportunities in the near or distant future to obtain the new observations required to model this better.

Evidence of a crater in the larger lobe previously discussed in Section \ref{subsec:planetaryradar} was also replicated despite the substantial penalties against concavities implemented to avoid over-fitting to the noise. A preliminary inspection of the model indicates the crater is $\sim 20$ m deep and $\sim 70$ m across. This crater could hint towards \DP's previous collisional or formation history; however, due to the limitations of the resolution of the model and the penalty functions used, no further conclusions can be reached at this time.

\begin{figure*}
	\includegraphics[width=\textwidth]{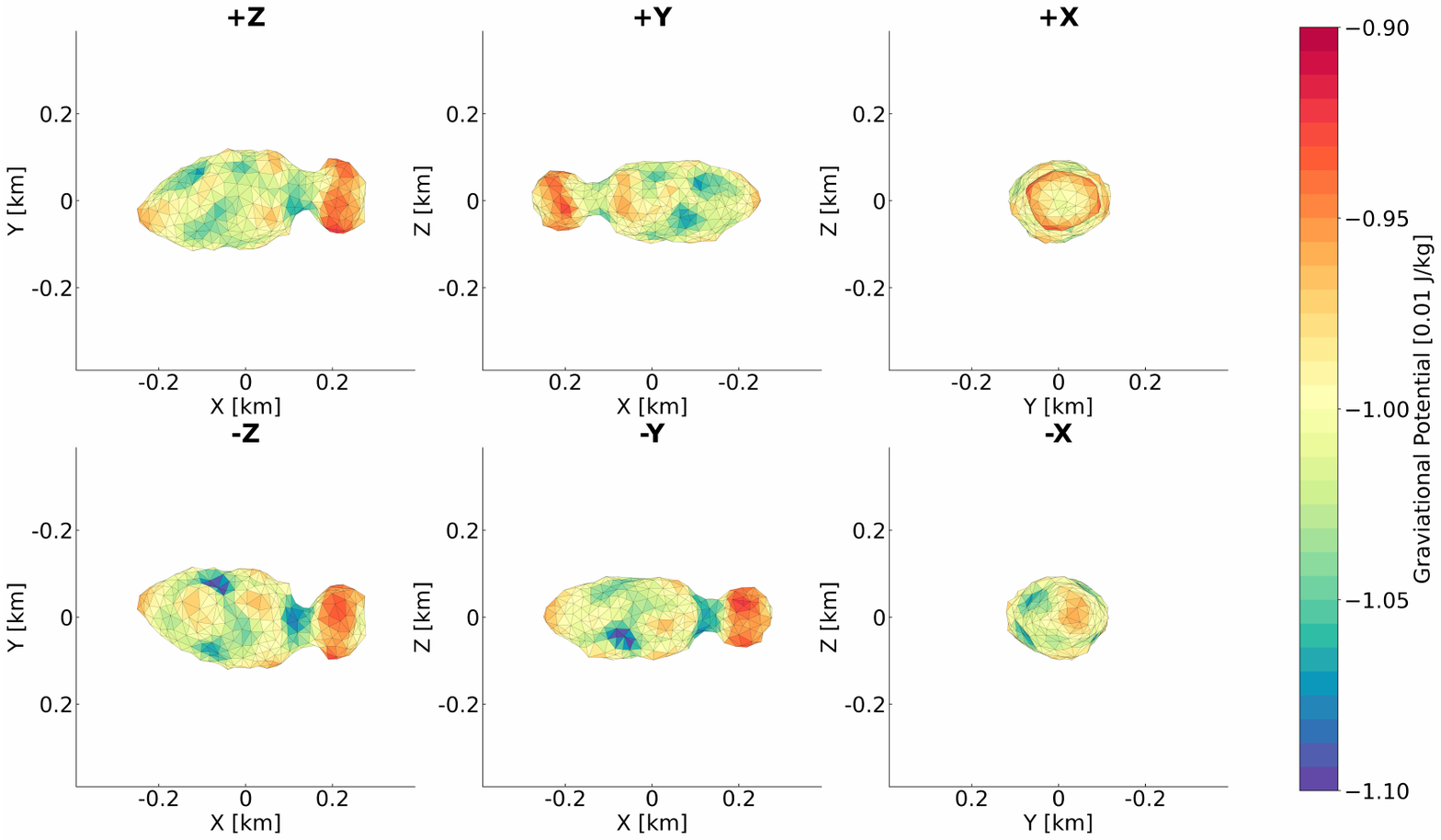}
    \caption{The gravitational potential across the surface of the best-fit model with pole solution $\lambda = 180^\circ$, $\beta = -80^\circ$ assuming a density of uniform $2035~\rm kg.m^{-3}$. The six plots show the views along the X, Y, and Z axes from both the positive and negative ends.}
    \label{fig:GravPotential}
\end{figure*}

\begin{figure*}
	\includegraphics[width=\textwidth]{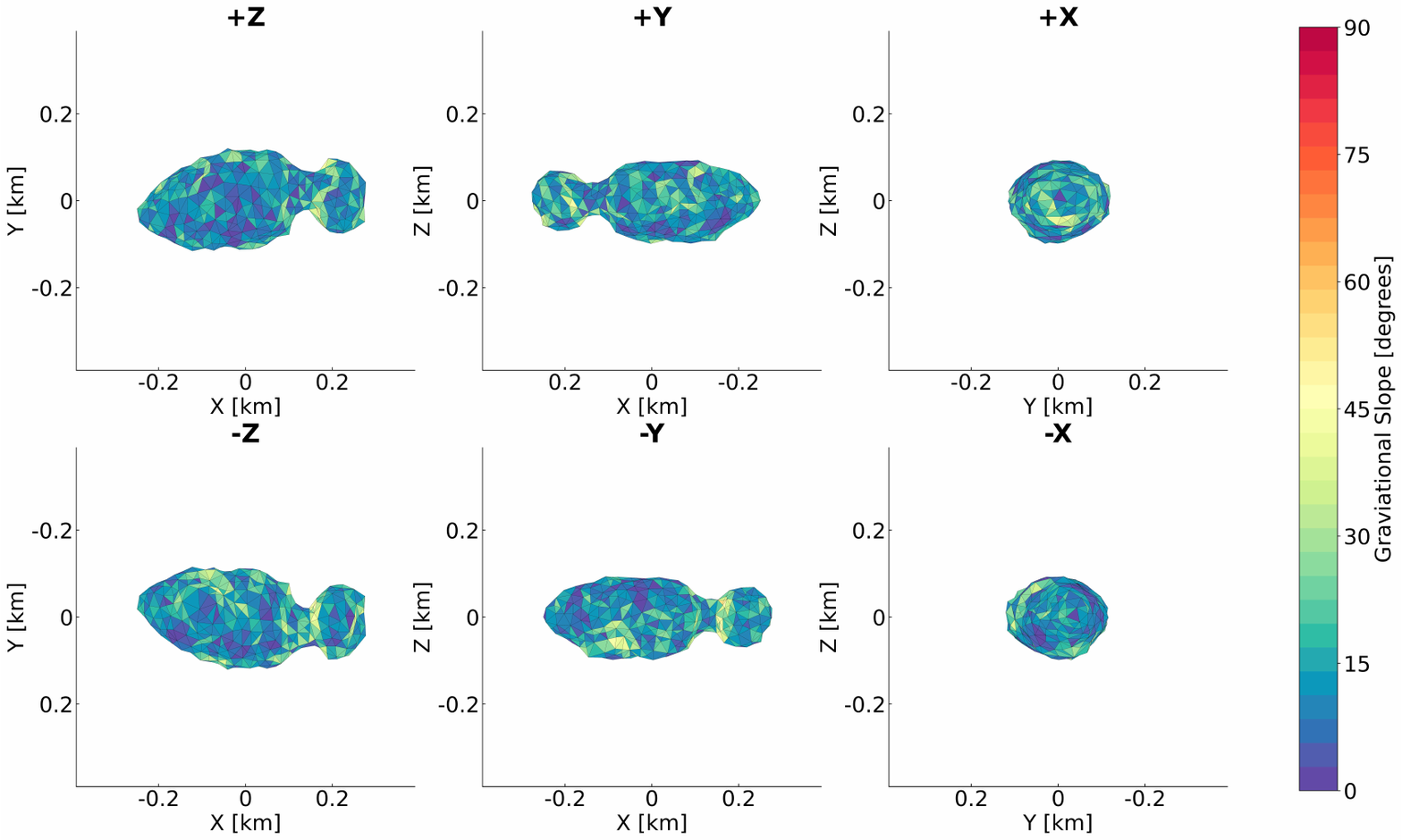}
    \caption{The gravitational slope of ambient gravity on the surface of the best-fit solution with rotational pole $\lambda = 180^\circ$, $\beta = -80^\circ$ assuming a uniform density of $2035~\rm kg.m^{-3}$. Ambient gravity is a measure that combines the gravitational force of the asteroid with the centrifugal force of the asteroid's rotation. The angle is measured relative to the normal of each facet. Therefore, any angle above 90 degrees is a force pointing away from the surface. The six plots show the views along the X, Y, and Z axes from both the positive and negative ends.}
    \label{fig:GravSlopes}
\end{figure*}

We can { constrain} the spectral classification of the asteroid using the radar albedo and the ratio $\rm SC/OC$ \citep{RiveraValentinEtAl2024_SpectralClassificationFromSCOC,VirkkiEtAl2014_SpectralClassificationFromSCOC+RadAlbedo}. 
In particular, these values { of $\rm SC/OC$ between 0.630 and 0.684} (Table~\ref{tab:AllRadarObs}) indicate that \DP~is likely an X, E{, or V-}type asteroid, which is in agreement with spectral analysis performed in 2014 by \citet{Hicks2014_DP14observations}{, which found \DP~to be an X- or C-type}. 
This is supported by an estimate of the optical albedo, defined as
\begin{equation}
    p_v = \left(10^{-0.2H}\frac{1329}{D_{eq}}\right),
\end{equation}
which, when using the MPC value of { H = }$19.0 \pm 0.5$ { and the calculated value of $D_{\rm eq} = 0.262\pm0.037$ km}, is calculated to be $p_v = (0.65\pm0.30)$. Therefore, \DP~is likely an E-type asteroid, as within the group of X-type asteroids, anything over $p_v > 0.3$ is likely to be an E-type \citep{ThomasEtAl2011_SpectralClassificationVsAlbedo,MahlkeEtAl2022_AsteroidSpectralTypeFromAlbedo}. However, we cannot use {the fact that \DP~is an E-type} to determine the density as X-complex asteroids are found to have a wide range of recorded densities \citep{Carry2012_DensityOfAsteroids}.

But what is the limiting density for which the lobes \DP~could stay together without any internal structure? We analysed the gravitational forces on the two lobes of \DP~by treating the lobes as two rigid uniform-density objects, as described in \citet{Scheeres2007_RotationalFissionOfCBs}. We find that a limiting uniform density of $\sim2035~\rm kg.m^{-3}$ is required in order to keep the critical spin period below $5.786$ hours and allow the lobes to be bound together under only the forces of gravity without the presence of any internal structural forces. Therefore, if \DP's density is less than this value, it would require internal cohesion for the lobes to remain connected. If \DP's density is more than this value, the lobes would be able to stay connected only under gravitational forces.

Proceeding with the limiting density of $2035~\rm kg.m^{-3}$, we calculated the gravitational environment across the surface of \DP. We find that the gravitational force is weakest on the smaller lobe and the end of the larger lobe and strongest around the centre of the larger lobe (Fig.~\ref{fig:GravPotential}). Additionally, the ambient gravity on the surface -- the combination of the gravitational forces and centrifugal forces acting on a point on the surface due to the asteroid's rotation - remains primarily perpendicular to the surface (Fig.~\ref{fig:GravSlopes}).

`Negative gravity', where the ambient gravity has a slope of $>90^{\circ}$ and points away from the surface, is a result of the centrifugal force generated by the asteroid's rotational speed overcoming the body's gravitational force. We find that such `negative' gravity only occurs on the surface of \DP~below densities of $1000~\rm kg.m^{-3}$, a regime where it is unusual to find C- or X-type asteroids. As there is no negative gravity, \DP~will be able to retain loose regolith and rocks on its surface, which is in line with current beliefs that most asteroids are `rubble piles' of loose rock and small boulders, a result of countless collisions since their formation \citep{JohansenEtAl2015-A4_Asteroid-Formation-Rubble-Piles}. The most significant gravitational slope on the surface of \DP~is $54^{\circ}$, with areas $>50^\circ$ present on the slopes of the connection between the neck and the smaller lobe and within the crater on the larger lobe. These regions may be regolith-free due to their slopes and demonstrate some internal cohesion. Consequently, this may imply a build-up of regolith at the narrowest point of the neck or the bottom of the crater.

{Finally, we can estimate the surface density of \DP~using the linear relation between the radar albedo and surface density introduced in \citet{OstroEtAl1985_DensityFromRadarAlbedo}:
\begin{equation}
    \rho = \frac{\hat{\sigma}_{OC}}{0.12g} + 1.803,
\end{equation}
where $g$ is the gain factor that accounts for the surface texture and shape (compared to an ideal sphere) of the object. We use the same value of $g = 1.2$ as \citet{ShepardEtAl2010_SurfaceDensityFromRadarAlbedo}, which analysed X and M-type asteroids in our calculations and found a surface density between approximately $2500$ to $3000~\rm kg.m^{-3}$. This range would suggest that {\DP~}would be gravitationally stable and not require internal cohesion. However, {surface density} is not equivalent to the bulk density of the object and cannot be used to assume the density below the surface.}

\subsection{Other contact binaries} \label{subsec:OtherCBs}

A current list of contact binaries, either observationally modelled or directly observed with spacecraft, is displayed in Table~\ref{tab:AllContactBinary} along with their physical properties, orbital classes and spectral types. While $D_{eq}$ (the diameter of an equivalent volume sphere) is commonly used to describe sizes of small solar system objects, for contact binaries which are commonly elongated, using the DEEVE, calculated using the technique described in \citet{Dobrovolskis1996_DEEVEcalc}, allows for more intuitive and useful comparisons. Notably, the DEEVE  parameters $2a$, $a/b$, and $b/c$ (where $a, b$, and $c$ are the semi-major axes of the ellipsoid and $a > b > c$) demonstrate a wide variety of morphologies can be classed under the title of `contact binary'. In particular, the two MBAs in Table~\ref{tab:AllContactBinary} are outliers compared to the NEA objects listed. (216) Kleopatra is over $300$ km long \citep{OstroEtAl2000_KleopatraInitialRadarModel, ShepardEtAl2018_KleopatraShapeModel}, $100$ times larger than the other objects, so likely formed through a very different mechanism where the gravitational forces play a much more significant role. On the other hand, Selam is the first ever contact binary found orbiting another body. As discussed in \citet{LevisonEtAl2024_LUCY-Dinkinesh-Characterisation-Selam-first-look}, it likely formed from mass shedding from (152830) Dinkinesh, the primary, and there is no evidence that the currently modelled NEA contact binaries were formed through the same process. As such, the 23 objects in Table~\ref{tab:AllContactBinary} cannot be treated as a cohesive population for study, emphasising the broad range of objects called `contact binaries' and the need for more contact binaries to be modelled such that they can be divided into groups based on their morphology and possible formation mechanisms.

A visual inspection of shape models for the NEAs in Table~\ref{tab:AllContactBinary} reveals differences in both the prominence of the neck structure and the sizes of the two lobes, which can be used to group the contact binaries.
Whilst the small sample size should be taken into consideration, a slight preference for unequally sized lobes is visible, with the larger lobe containing more than 66\% of the total volume in 60\% of cases.
{The inequality between the sizes of \DP's lobes places it as one of these such objects. 
Other NEA contact binaries with similar shape} include (25143) Itokawa~\citep{DemuraEtAl2006_Itokawa-Shape}, (85990) 1999 \JV~\citep{RozekEtAl2019_JV6-shape}, (85989) 1999 \JD~\citep{Marshall2017_PhDthesis_JD6_shape}, and (8567) 1996 \HW~\citep{MagriEtAl2011_1996HW1-shape-CB} (henceforth \JV, \JD, and \HW), {while more symmetrical objects include} Selam~\citep{LevisonEtAl2024_LUCY-Dinkinesh-Characterisation-Selam-first-look}, or (4769) Castalia \citep{Hudson&Ostro1994_Castalia-Shape}.
Within the above selection of contact binaries with unequally distributed mass, \DP~shares closer morphology with \JD~and \HW~due to the distinct narrow neck connecting the lobes — in contrast to Itokawa and \JV, where the lobes appear as two overlapping ellipsoids. Clearly defined necks in contact binaries may indicate a gentle collision or the presence of structural rigidity, compared to the thicker necks that may indicate the two lobes were deformed when they came together. \DP's larger lobe contains 84\% of the asteroid's mass (assuming uniform density), which is high compared to \HW's 66\% or \JV's 70\%, but very similar to Itokawa's 84\%, despite the differing neck morphology. For reference, Castalia's larger lobe contains 60\% of its total mass. It should be noted, however, that \HW~and \JD~are over $4$ and $3$ km long compared to the $0.5$ km length of \DP, and all three asteroids are of different spectral classes. Therefore, despite their shared morphology, we cannot state that they may have similar formation histories.

The crater that appears on \DP's larger lobe has similarities to a number of other modelled contact binaries, with several other modelled contact binaries having notable concavities on their surface. Examples of these include \HW, (11066) Sigurd, (4179) Toutatis, and (486958) Arrokoth. Work has been done in analysing the effects that an impact crater would have on a contact binary if one assumes the impact occurs once the object already has a contact binary shape \citep{Hirabayashi2023_CraterImpactOnContactBinarys}, especially with regards to Arrokoth. \citet{HirabayashiEtAl2020_ArrokothCraterAndShapeChange} finds that it is possible for an object to reform into a contact binary structure even if an impact were to break the neck structure of an initial bi-lobed shape, as long as the lobes remain tidally locked to each other. Alternatively, if the object had enough cohesive strength, the lobes could stay connected, and the impact would only alter the spin state of the object. Therefore, it cannot be determined whether \DP's crater formed before or after its contact binary structure was developed.

As mentioned in Section \ref{sec:introduction}, while a single detailed model does not contain enough information to infer how the object is likely to have formed, by creating more shape models, we hope to open the door in the future for analysis of multiple objects at a time. Currently, 23 contact binaries have been modelled, and the term covers more elongated objects like Kleopatra \citep{ShepardEtAl2018_KleopatraShapeModel}, equally sized lobes such as Selam \citep{LevisonEtAl2024_LUCY-Dinkinesh-Characterisation-Selam-first-look}, and the bi-lobed objects with uneven mass distributions like Itokawa \citep{DemuraEtAl2006_Itokawa-Shape} and \DP. Additionally, the range of sizes covered by this term spans from the 310 km long oddball Kleopatra to 500 m long or shorter. The longest contact binary NEAs are all in the region of $4$ km long. This provides a size range of at least an order of magnitude within the contact binary NEA population. By increasing the number of similarly shaped objects, an analysis could be performed on groups with similar shapes to investigate whether they are likely to have formed through the same mechanism or if multiple formation pathways result in the same type of morphology. These possible formation pathways are described in Section~\ref{subsec:CBformation}.

\begin{landscape}
    \begin{table}
        \centering
        \caption{For each modelled contact binary, the following properties are given: the equivalent diameter (diameter of equal volume sphere), the rotational period, three DEEVE parameters, the spectral type (if known/estimated) and whether it was observed by occultations (O), lightcurves (LC), radar (R), or direct spacecraft observations (S). The class of object is also given (NEA -- near-Earth asteroid, MBA -- main-belt asteroid, KBO -- Kuiper Belt object, HTC -- Halley-type comet, and SPC -- short period comet).
        Filled bullet points ($\bullet$) signify techniques directly used in the paper(s) cited. Hollow bullet points ($\circ$) represent techniques used for less detailed models/spin state analysis which agree with the cited model.
        Where provided, uncertainties in this table are taken or calculated directly from the referenced material and are often derived using different methods for different studies.
        DEEVE parameters were either taken from the referenced material or calculated from publicly available vertex models, and as such, errors are not provided for these values but can be found in the referenced material for a selection of the objects.
        Models supplied by \privcomtBusch~ do not have errors associated with them due to still being in an early stage of modelling. 
        {$\ast$} Selam, 8P/Tuttle, and 19P/Borelly have no complete vertex model published. The values provided for $2a$ are not DEEVE values but are estimates of length based on initial images from the LUCY mission, radar observations, and DS1 flyby, respectively. 
        {$\dagger$} Toutatis and Apophis are tumbling (non-principal axis rotators). In this case, the periods given are the best stand-in to fit a single rotation to the lightcurve.
        {$\ddagger$} Arrokoths shape was previously accepted to be a flatter shape \citep{KeaneEtAl2022_Arrokoth-Shape-Flyby}; however, newer analysis of New Horizons data \citep{PorterEtAl2024_ConfAbs_ArrokothUpdatedShape} suggest a thicker shape, which is the model described in this table. The period given here, however, is taken from \citet{BuieEtAl2020_ArrokothPeriod}. 
        {$\star$} The period of comet 103P was changing quickly while approached by the EPOXI mission. This 16 hour period is the most relevant to the model \citep{ThomasEtAl_103P-spacecraft-images-shape}.
        \textbf{References:} 
        (1) \citet{McGlassonEtAl2022_MidasShapeModel}
        (2) \citet{BennerEtAl1999_BacchusShapeModel}
        (3) \citet{Hudson&Ostro1994_Castalia-Shape}
        (4) \citet{HudsonEtAl2003_Toutatis-shape-CB}
        (5) \citet{ZhaoEtAl2016_Toutatis-Shape-Radar-Spacecraft}
        (6) \citet{HowellEtAl1994_ToutatisSpectralClass}
        (7) \citet{BrozovicEtAl2010_Mithra-Shape-Model}
        (8) \citet{BinzelEtAl2019_SpectralClassSurvey}
        (9) \citet{MagriEtAl2011_1996HW1-shape-CB}
        (10) \privcomtBusch ~
        (11) \citet{Fevig&Fink2007_SigurdSpectralClass}
        (12) \citet{PravecEtAl1998_SigurdPeriod}
        (13) \citet{BennerEtAl2004_SigurdRadarImaging}
        (14) \citet{DemuraEtAl2006_Itokawa-Shape}
        (15) \citet{GaskellEtAl2008_ItokawaVertexModel}
        (16) \citet{LedererEtAl2005_ItokawaSpectraClass}
        (17) \citet{ZegmottEtAl2021_KZ66-YORP-Detection-and-Shape}
        (18) \citet{Marshall2017_PhDthesis_JD6_shape}
        (19) \privcomtMarshall ~
        (20) \citet{RozekEtAl2019_JV6-shape}
        (21) \citet{BrozovicEtAl2018_ApophisRadarObsAndModel}
        (22) \citet{PravecEtAl2014_ApophisPeriodTumbling}
        (23) \citet{BinzelEtAl2009_ApophisSpectralClass}
        (24) \citet{BrauerEtAl2014_ConfAbs2000rs11_shape}
        (25) \citet{BinzelEtAl2004_SpectralClassSurvey}
        (26) \citet{ShepardEtAl2018_KleopatraShapeModel}
        (27) \citet{LevisonEtAl2024_LUCY-Dinkinesh-Characterisation-Selam-first-look}
        (28) \citet{deLeonEtAl2023_DinkineshSelamSpectralClass}
        (29) \citet{PorterEtAl2024_ConfAbs_ArrokothUpdatedShape}
        (30) \citet{BuieEtAl2020_ArrokothPeriod}
        (31) \citet{GroussinEtAl2019_8P-Shape-Spitzer-Obs}
        (32) \citet{HarmonEtAl2010_8P_radar_paper}
        (33) \citet{OberstEtAl2004_19P-spacecraft-images-shape}
        (34) \citet{SierksEtAl2015_67P-Shape}
        (35) \citet{GaskellEtAl2017_67PVertexModel}
        (36) \citet{ThomasEtAl_103P-spacecraft-images-shape}
        (37) \citet{FarnhamEtAl2013_103PvertexModel}
        }
        \label{tab:AllContactBinary}
        \begin{tabular}{lcccccccccccc} \hline \hline 
        Object                              & Class     & $D_{eq}$                 & Period                                & DEEVE     & DEEVE     & DEEVE     & Spectral  & O         & LC        & R         & S         & Ref.  \\ 
        Name                                &           & [km]                      & [h]                                   & $2a$ [km] & $a/b$     & $b/c$     & type      &           &           &           &           &       \\ \hline
        (1981) Midas                        & NEA       & $1.77 \pm 0.26$           & $5.2216\pm 0.0017$                    & 3.59      & 2.36      & 1.51      & V         &           & $\bullet$ & $\bullet$ &           & (1)   \\
        (2063) Bacchus (2 lobes)            & NEA       & $0.67^{+0.13}_{-0.07}$    & $15^{+0.2}_{-0.2}$                    & 1.12      & 2.30      & 1.03      & Q / C     &           & $\circ$   & $\bullet$ &           & (2)   \\
        (4769) Castalia                     & NEA       & 1.08                      & $4.07\pm 0.02$                        & 1.73      & 1.86      & 1.18      & --        &           &           & $\bullet$ &           & (3)   \\
        (4179) Toutatis {$\dagger$}         & NEA       & 2.44                      & $175$                                 & 4.25      & 2.09      & 1.09      & S         &           & $\circ$   & $\bullet$ & $\bullet$ & (4)(5)(6)   \\
        (4486) Mithra (Prograde)            & NEA       & $1.69 \pm 0.05$           & $67.5\pm 6.0$~                        & 2.56      & 1.70      & 1.21      & S         &           &           & $\bullet$ &           & (7)(8)   \\
        \hspace{1.66cm}(Retrograde)         & NEA       & $1.75 \pm 0.05$           & $67.6\pm 6.0$~                        & 2.66      & 1.72      & 1.20      & S         &           &           & $\bullet$ &           & (7)(8)   \\
        (8567) 1996 HW$_{\rm 1}$            & NEA       & $2.02 \pm 0.16$           & $8.76243\pm 0.00004$                  & 4.20      & 2.84      & 1.11      & Sx        &           & $\bullet$ & $\bullet$ &           & (9)   \\
        (11066) Sigurd                      & NEA       & $2.14 \pm 0.34$           & $8.4958 \pm 0.0003$                   & 4.46      & 2.89      & 1.08      & S         &           & $\circ$   & $\bullet$ &           & (10)(11)(12)(13) \\
        (25143) Itokawa                     & NEA       & $0.324 \pm 0.006$         & $12.132371 \pm 0.000006$              & 0.59      & 2.25      & 1.19      & S         &           & $\circ$   & $\circ$   & $\bullet$ & (14)(15)(16)   \\
        (68346) 2001 KZ$_{\rm 66}$          & NEA       & 0.797                     & $4.985997\pm 0.000042$                & 1.58      & 2.51      & 1.24      & S         &           & $\bullet$ & $\bullet$ &           & (17)   \\
        (85989) 1999 JD$_{\rm 6}$           & NEA       & $1.45 \pm 0.14$           &$7.6643464_{-0.0000058}^{+0.0000054}$  & 3.17      & 2.19      & 1.16      & K         &           & $\bullet$ & $\bullet$ &           & (18)(19)  \\
        (85990) 1999 JV$_{\rm 6}$           & NEA       & $0.442\pm0.004$           & $6.536787\pm 0.000007$                & 0.78      & 2.27      & 1.04      & Xk        &           & $\bullet$ & $\bullet$ &           & (20)  \\
        (99942) Apophis  {$\dagger$}        & NEA       & $0.34 \pm 0.04$           & $30.56\pm 0.01$~                      & 0.40      & 1.25      & 1.07      & Sq        &           & $\circ$   & $\bullet$ &           & (21)(22)(23)    \\
        (275677) 2000 RS$_{\rm 11}$         & NEA       & $3.50 \pm 0.35$           & $4.444\pm 0.001$                      & 4.04      & 1.04      & 1.41      & --        &           & $\circ$   & $\bullet$ &           & (24) \\
        (388188) 2006 \DP~                  & NEA       & $0.262 \pm 0.037$         & $5.7860\pm0.0001$                     & 0.52      & 2.57      & 1.19      & E         &           & $\bullet$ & $\bullet$ &           & This work \\
        (523804) 2000 YF$_{\rm 29}$         & NEA       & $0.30 \pm 0.04$           & $6.14 \pm 0.15$                       & 0.42      & 1.49      & 1.26      & S         &           &           & $\bullet$ &           & (10)(25) \\
        2004 XL$_{\rm 14}$                  & NEA       & $0.21 \pm 0.03$           & $17.0 \pm 0.4$~                       & 0.30      & 1.40      & 1.52      & C         &           &           & $\bullet$ &           & (8)(10) \\ \hline
        (216) Kleopatra                     & MBA       & $122 \pm 30$~             & $5.385280\pm 0.000001$                & 310       & 3.67      & 1.21      & M         & $\bullet$ & $\circ$   & $\bullet$ &           & (26)   \\
        (152830) Dinkinesh I Selam $\ast$   & MBA       & --                        & $52.44\pm 0.14$                       & 0.44      & --        & --        & S         &           &           &           & $\bullet$ & (27)(28)  \\ \hline
        (486958) Arrokoth~~{$\ddagger$}     & KBO       & 19.9                      & $15.9380\pm 0.0005$                   & 38.0      & 2.28      & 1.34      &           & $\circ$   & $\circ$   &           & $\bullet$ & (29)(30)  \\
        8P/Tuttle $\ast$                    & HTC       & --                        & 11.4                                  & 10        & --        & --        &           &           & $\bullet$ & $\bullet$ &           & (31)(32)  \\
        19P/Borrelly $\ast$                 & SPC       & --                        & --                                    & 7.5       & --        & --        &           &           &           &           & $\bullet$ & (33)  \\
        67P/Churyumov–Gerasimenko           & SPC       & 3.30                      & $12.4043\pm 0.0007$                   & 4.68      & 1.57      & 1.15      &           &           &           &           & $\bullet$ & (34)(35)  \\
        103P/Hartley 2  {$\star$}           & SPC       & 1.15                      & 16                                    & 2.57      & 3.11      & 1.13      &           &           &           & $\circ$   & $\bullet$ & (36)(37)  \\ \hline
        \end{tabular}
    \end{table}
\end{landscape}

\subsection{Contact Binary Formation} \label{subsec:CBformation}

The most obvious explanation for two objects to combine and stay together is for them to approach each other slowly and stick together under gravitational forces and the presence of `sticky' materials such as volatile ices. Indeed, models show that Kuiper belt objects, such as the contact binary Arrokoth, could form contact binaries in collisions, as long as the energy is on the order of the object's escape velocity or less \citep{Jutzi&Asphaug2015_Low-Velocity-CB-formation-comets}. This may also explain the high number of contact binary candidates in the KBO and cometary populations. Whilst this explanation is ideal for icier objects in the outer solar system, the inner solar system has higher temperatures that will cause outgassing of ices from the surface \citep{Schorghofer2018_AsteroidsOutgassingIces}. Although ice could be present in the larger NEAs if they were transported from the outer main belt \citep{SchorghoferEtAl2020_Ice-in-NEOs-from-Outer-MBA}, the outgassing timescales for an object the size of \DP, if placed stationary at its aphelion, would be $<10^4$ years. For \DP~specifically, its eccentric orbit and its rubble pile nature would have accelerated this process beyond this theoretical value. As most NEAs are considered to have a rubble pile constitution, the formation of contact binaries in the inner solar system requires explanations without the presence of volatile ices and higher relative velocities.

One explanation is that bi-lobed objects in higher energy environments form from the re-accumulation of fragments following a catastrophic collision. \citet{CampoBagatinEtAl2020_CB-Formation-Collisions-Itokawa-Sim} and \citet{Michel&Richardson2013_Grav-Reaccumulation-Itokawa} both attempted to simulate such an occurrence to replicate the shape of Itokawa and found that, under the right conditions, an approximate shape could be reached. \citet{CampoBagatinEtAl2020_CB-Formation-Collisions-Itokawa-Sim} found that a fragmented body could re-accumulate by itself into a contact binary structure, as collapsing fragments could `bump' the largest fragment away from the centre of mass, with it forming the head of a contact binary. It should be noted that there are other theories for Itokawa's formation - \citet{LowryEtAl2014_Itokawa-YORP-and-densities} found that even with a highly constrained shape and spin state model and gravitational measurements from Hayabusa, the mass could either be distributed unevenly such that the head and body have different densities - perhaps implying different parent bodies - or the same density with a higher density region in the neck: possibly compressed when two objects slowly combined.

There is also a mechanism for two objects close to the Sun to merge in a gentle collision. The binary YORP (BYORP) effect is a variation of the Yarkovsky and YORP effects -- effects caused by the asymmetric re-emission of thermal radiation absorbed from the Sun of a rotating body with some thermal inertia -- that affects the orbital dynamics of a binary asteroid system \citep{BottkeEtAl2001_Yarkovsky-spreading-families, Rubincam2000_YORP-concept, Cuk&Burns2005_BYORP-concept, VokrouhlickyEtAl2015-A4_Yarkovsky-YORP-Review}. These forces can combine in such a way as to cause the secondary asteroid to either slowly in-spiral or out-spiral. If the system is doubly synchronous (both objects tidally locked to one another), then a slow in-spiral could form a contact binary \citep{Jacobson&Scheeres2011_CB-formation-from-BYORP}. However, calculations find that slowly in-spiralling bodies may also be affected by tidal forces between the asteroids, resulting in an equilibrium state between BYORP and tidal forces that stops any change in orbital properties \citep{Jacobson&Scheeres2011_Tidal-eq-BYORP}.

With the recent discovery of Selam by the LUCY mission, there is now a new theory. Selam is the first contact binary that has been found orbiting another object, and so questions arise around whether contact binaries could form as secondaries in multiple object systems before becoming separated from their primary, perhaps from an out-spiralling BYORP \citep{LevisonEtAl2024_LUCY-Dinkinesh-Characterisation-Selam-first-look}. Multiple mass wasting events from the primary, or a single event that produces a debris disk around the object, could then re-accumulate into a contact binary structure \citep{WimarssonEtAl2024_DebrisDiskContactBinaryFormation}.

Even for the most well-observed contact binary, Itokawa, with extensive ground-based observations in multiple wavelengths and detailed spacecraft observations with the Hayabusa mission, we cannot definitively say which formation pathway the asteroid took.
{ Simulations of asteroid formation can provide further context for contact binary formation methods, especially for objects closer to an ellipsoid. Still, they can struggle to replicate thin neck-like structures seen on \DP~or \JD~at current resolutions. Therefore, it is important to increase both the quality and quantity of both simulated and observed models in order to provide a larger and more diverse sample of objects for comparison between the two populations.}

\subsection{Future observations} \label{subsec:FutureObservations}

\DP's following close approaches in 2030 and 2031, while not approaching close enough for high-quality radar observations (0.29 au in 2030 and 0.19 au in 2031), will be good opportunities to gather more optical data to constrain the spin state further. More lightcurves 7 years later would allow a thorough investigation into any YORP and Yarkovsky effects and allow a stronger estimate for the period and pole solution. The 2030 observing epoch, in particular, would be incredibly useful due to the viewing angles of $\alpha = 145^\circ$, $\lambda = 261^\circ$, and $\beta = 6^\circ$, offering an entirely new viewing geometry compared to our current data set (Table~\ref{tab:AllLightcurves}. The 2031 epoch would have similar phase angles to the 2023 epoch, which would also be useful as this is currently the noisiest dataset for which we have the fewest lightcurves. Further optical observations would also be able to obtain a more detailed spectral analysis to refine the spectral classification of \DP. Without significant improvements in radar technology, the next opportunity to observe \DP~with radar will be in 2065. However, as the closest approach would only be $\sim0.05$ au, only CW observations could likely be collected, as SNR for delay-Doppler images would be too low to create useful images. Without improvement in Goldstone-equivalent radar facilities, \DP~will not come close enough to Earth for detailed delay-Doppler imaging again until at least 2195.

\section{Conclusions} \label{sec:Conclusions}

It is estimated from radar observations that 15-30\% of NEAs are bi-lobed in shape \citep{BennerEtAl2015A4_15-percent-CB-NEO,VirkkiEtAl2022_30percent-CB}. The current selection of observed contact binaries demonstrates a wide variety of morphologies with varying distributions of masses between the lobes and neck structures, which likely form through many different formation mechanisms. For example, Selam, the first ever contact binary moon, likely formed in a different manner than Itokawa or other lone contact binaries \citep{LevisonEtAl2024_LUCY-Dinkinesh-Characterisation-Selam-first-look}.

Only by modelling more contact binaries can these objects be placed in context with one another and sub-characterised. Already, a visual inspection reveals similarities between several modelled objects, such as the group of objects with a smaller lobe attached to a larger elongated ellipsoid, similar to \DP.
{Indeed, approximately 60\% of the 16 modelled NEA contact binaries discussed in this paper have a mass distribution between lobes of 2:1 or greater.}

The final model for \DP~has a period of $(5.7860\pm0.0001)$ hours with pole solution of $\lambda = (180\pm121)^\circ$ and $\beta = (-80\pm7)^\circ$. The shape solution created by combining radar and optical observations shows a 500-meter-long object of bi-lobed structure. The larger lobe, contributing 84\% of the volume, is elliptical in shape and features a large crater $\sim20$ m deep and $\sim70$ m across. The larger lobe connects to the more spherical smaller lobe at one end through a narrow neck, which has a radius of $39$ metres.

The presented model is at the resolution of the radar observations, and a more detailed model would not be possible with the current data set without the risk of over-fitting noise. Future observations with optical telescopes may improve the estimation of the spin state, particularly the pole solution. However, the earliest these will be able to be collected is 2030, while radar observations could be collected in 2065.

\section*{Acknowledgements}
%

%
We acknowledge and thank the late Joseph Pollock for his contribution to astronomy and key role in collecting two of the lightcurves used in this analysis.
We thank all the technical and support staff at the observatories at La Silla, Cerro Tololo, Perth Observatory, and Roque de los Muchachos for their support and the staff at Goldstone for their help with the radar observations. This material is based in part upon work supported by NASA under the Science Mission Directorate Research and Analysis Programs.
REC, AR, CS and AD acknowledge the support from the UK Science and Technology Facilities Council.
Part of this work was carried out at the Jet Propulsion Laboratory, California Institute of Technology, under contract with the National Aeronautics and Space Administration (80NM0018D0004).
The work at Ond\v{r}ejov was supported by {\it  Praemium Academiae} award to P. Pravec by the Academy of Sciences of the Czech Republic.
TH acknowledges funding from the Public Scholarship, Development, Disability and Maintenance Fund of the Republic of Slovenia.
UGJ acknowledges funding from the Novo Nordisk Foundation Interdisciplinary Synergy Programme grant no. NNF19OC0057374.
EK is supported by the National Research Foundation of Korea 2021M3F7A1082056.
PLP was partly funded by the FONDECYT Initiation Project No. 11241572. R.F.J. acknowledges support for this project provided by ANID's Millennium Science Initiative through grant ICN12\textunderscore 009, awarded to the Millennium Institute of Astrophysics (MAS), and by ANID's Basal project FB210003.
For the purpose of open access, the author has applied a Creative Commons Attribution (CC BY) licence to any Author Accepted Manuscript version arising from this submission.
This is a pre-copyedited, author-produced PDF of an article accepted for publication in MNRAS following peer review.

%
%
%
%
\section*{Data Availability}

The shape model described in this report is available online and for download from \href{https:/3d-asteroids.space}{https://3d-asteroids.space}. Lightcurve IDs 3 and 4 are available on \href{https://alcdef.org/}{ALCDEF}.
The light curve data first published here can be accessed at the CDS via anonymous ftp to cdsarc.u-strasbg.fr (130.79.128.5) or via \url{http://cdsarc.u-strasbg.fr/viz-bin/qcat?J/MNRAS}.
Radar data is available upon request but will be in the future made available through the PDS Asteroids/Dust Subnode: \url{https://sbn.psi.edu/pds/archive/asteroids.html}.




\bibliographystyle{mnras}
\bibliography{MNRAS_388188,custom_refs}




\appendix

\section{Convex Inversion results} \label{app:ConvInvResults}

\begin{figure*}
	\includegraphics[width=0.95\textwidth]{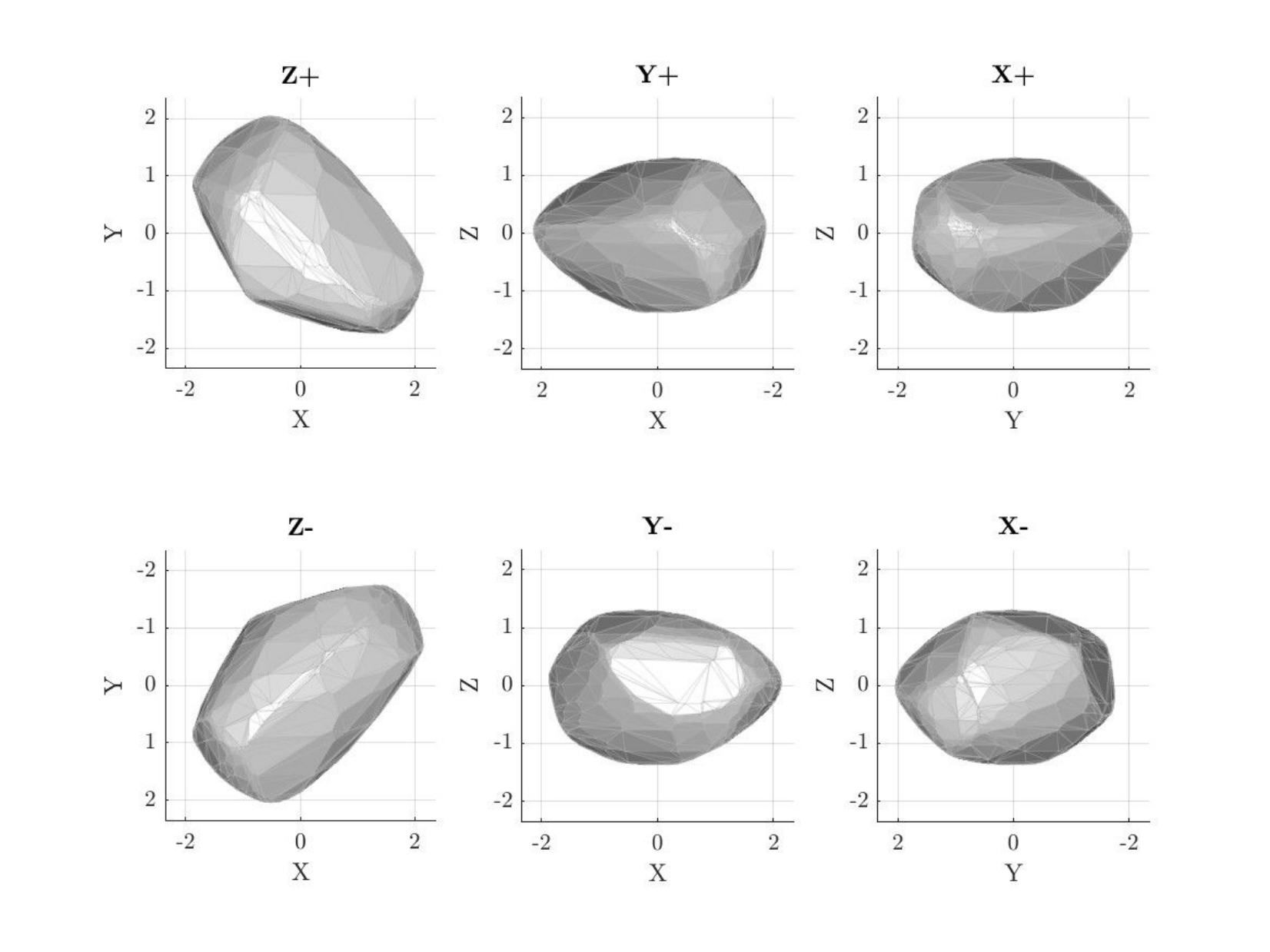}
    \caption{The best fitting model of \DP~from the convex inversion modelling, with pole solution $\lambda = 235$, $\beta = -65$. The six plots show the views along the X, Y, and Z axes from both the positive and negative ends.}
    \label{figapp:M1_model}
\end{figure*}

\begin{figure*}
	\includegraphics[width=0.95\textwidth]{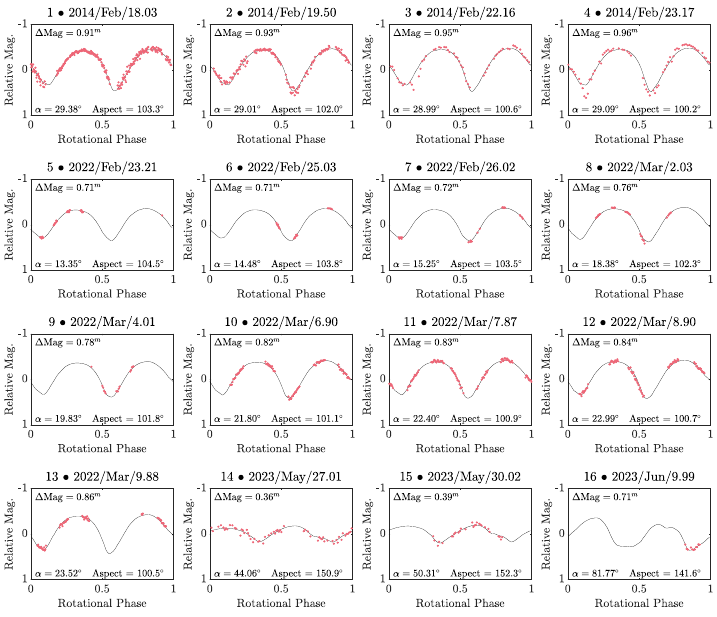}
    \caption{The lightcurve fits for the convex inversion model of \DP~with pole solution $\lambda = 235$, $\beta = -65$. The red dots are the provided data, and the black line is the simulated lightcurve of the model, assuming Lommel-Seelinger scattering. {Listed on the figure is $\Delta \rm Mag$, the peak-to-peak magnitude of the lightcurve, $\alpha$, the mean solar phase angle for the duration of the lightcurve, and the mean aspect angle over the duration of the lightcurve}}  
    \label{figapp:M1_lcfit}
\end{figure*}

\section{Radar modelling results} \label{app:RadarModellingResults}

\begin{figure*}
	\includegraphics[width=0.75\textwidth]{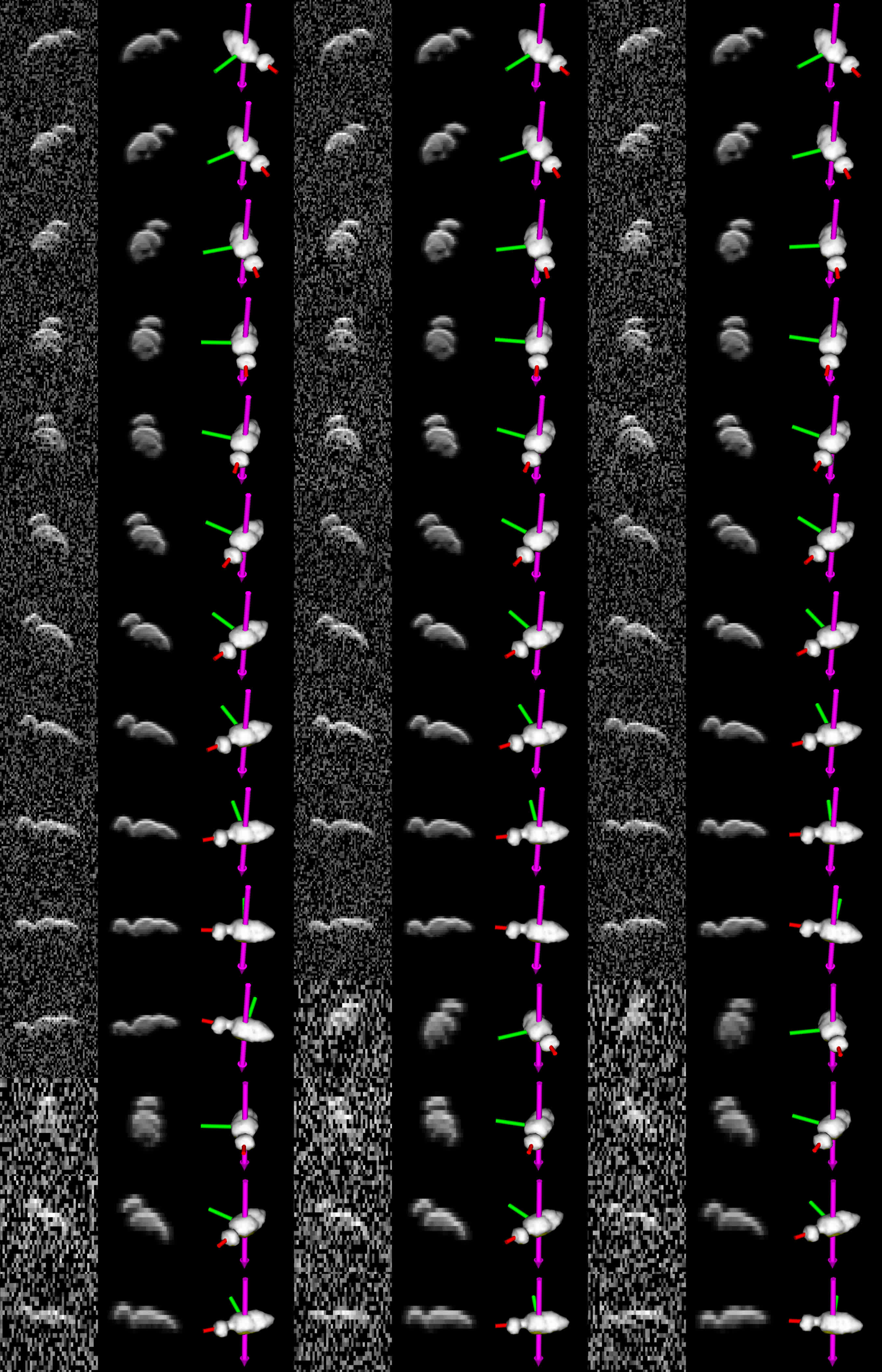}
    \caption{The raw and simulated delay-Doppler images for the SHAPE model with pole solution $\lambda = 180$, $\beta = -80$. Snapshots are displayed in 3 columns, in chronological order, read left to right, each consisting of 3 images. \textit{Left}: The raw delay-Doppler images. \textit{Centre}: The simulated delay-Doppler images for the model at that time. \textit{Right}: The plane-of-sky view of the model, as it would be seen by direct optical imaging.}  
    \label{figapp:rad_ddfit1}
\end{figure*}

\begin{figure*}
	\includegraphics[width=0.95\textwidth]{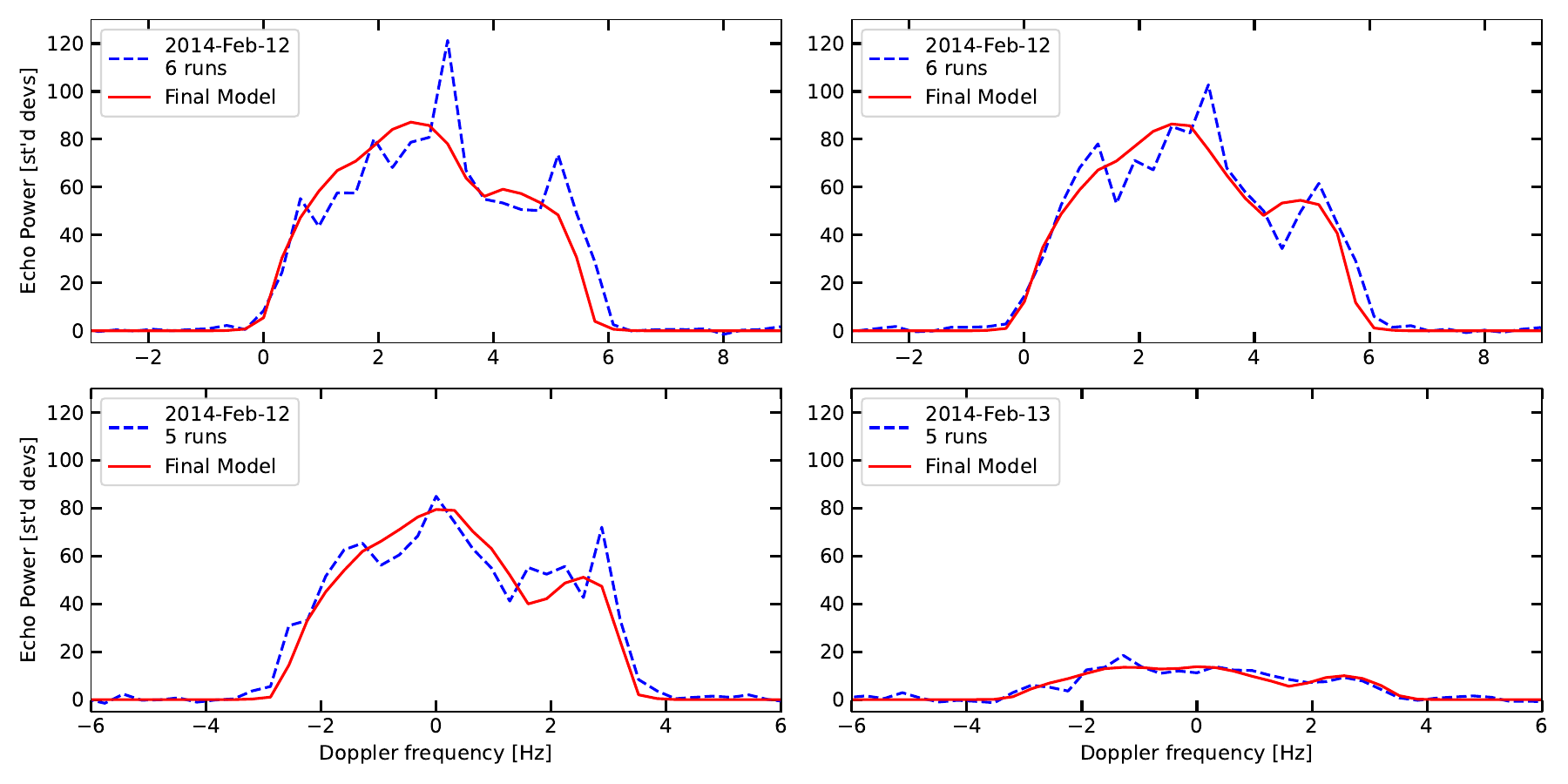}
    \caption{The continuous wave power spectra data used in the modelling (dashed blue line) and the simulated power spectra from the best-fit model (red line). Power spectra were stacked to increase SNR, with data from 12$^{\rm th}$ of February being split into two groups of 6 runs and one group of 5. Data on the 13$^{\rm th}$ of February was significantly weaker and stacked in a group of 5 runs. The spectral resolution of the data across all 3 days was $0.32$ Hz.}  
    \label{figapp:rad_cwfit}
\end{figure*}

\begin{figure*}
	\includegraphics[width=0.98\textwidth]{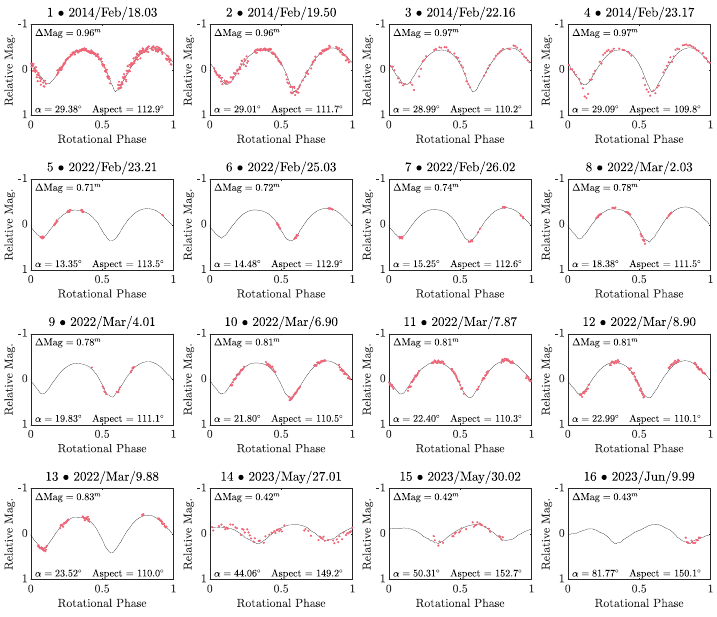}
    \caption{The lightcurve fits for the radar SHAPE model of \DP~with pole solution $\lambda = 180$, $\beta = -80$. The red dots are the provided data, and the black line is the simulated lightcurve of the model, assuming Lommel-Seelinger scattering. {Listed on the figure is $\Delta \rm Mag$, the peak-to-peak magnitude of the lightcurve, $\alpha$, the mean solar phase angle for the duration of the lightcurve, and the mean aspect angle over the duration of the lightcurve}}  
    \label{figapp:rad_lcfit}
\end{figure*}


\bsp	
\label{lastpage}
\end{document}